%% file: delabrouille-diffuse-nopdf.tex
\documentclass[multphys,vecphys]{svmult}

\usepackage[latin1]{inputenc} 
\usepackage{makeidx}         
\usepackage{url}         
\usepackage{graphicx}        
\usepackage{multicol}        
\usepackage[bottom]{footmisc}

\makeindex             

\newcommand\mypar[1]{\paragraph*{#1}}
\newcommand{\m}[1]{\bf{#1}} 
\newcommand{\mA}{{\bf{A}}} 
\newcommand{\mRs}{{\bf{R_s}}} 
\newcommand{\mRn}{{\bf{R_n}}} 
\newcommand{\mRx}{{\bf{R_x}}} 
\newcommand{\mRy}{{\bf{R_y}}} 
\newcommand{\mRone}{{\bf{R_1}}} 
\newcommand{\mRtwo}{{\bf{R_2}}} 
\newcommand{\mC}{{\bf{C}}} 
\newcommand{\mW}{{\bf{W}}} 
\newcommand{\mM}{{\bf{M}}} 
\newcommand{\covar}{{\rm{cov}}}

\def\A{\mathbf{A}}
\def\R{\mathbf{R}}
\def\W{\mathbf{W}}
\def\N{\mathbf{N}}
\def\Sig{\mathbf{\Sigma}}

\def\bs{\mathbf{s}}
\def\bx{\mathbf{x}}
\def\by{\mathbf{y}}
\def\bn{\mathbf{n}}

\def\inv{^{-1}}
\def\adj{^\dagger}

\def\nbs{n}
\def\nbe{P}

\begin{document}

\title*{Diffuse source separation in CMB observations}
\author{J. Delabrouille \inst{1}\and
J.-F. Cardoso \inst{2}}
\institute{APC, 11 place Marcelin Berthelot, F75231 Paris Cedex 05 \\
\texttt{delabrouille@apc.univ-paris7.fr}
\and LTCI, 46 rue Barrault, F75634 Paris Cedex 13 and \\ APC, 11 place Marcelin Berthelot, F75231 Paris Cedex 05 \\
\texttt{cardoso@enst.fr}}
%
%
\maketitle

\section{Introduction}

Spectacular advances in the understanding of the Big-Bang model of cosmology have been due to increasingly accurate observations of the properties of the Cosmic Microwave Background (CMB). The detector sensitivities of modern experiments have permitted to measure fluctuations of the CMB temperature with such a sensitivity that the contamination of the data by astrophysical foreground radiations, rather than by instrumental noise, is becoming the major source of limitation.
This will be the case, in particular, for the upcoming observations by the Planck mission, to be launched by ESA in 2008 \cite{2003NewAR..47.1017L,2000ApL&C..37..151M,2000ApL&C..37..161L}, as well as for next generation instruments dedicated to the observation of CMB polarisation.

In this context, the development of data analysis methods dedicated to identifying and separating foreground contamination from CMB observations is of the utmost importance for future CMB experiments. In many astrophysical observations indeed, and in particular in the context of CMB experiments, signals and images contain contributions from several components or {\emph{sources}}. Some of these sources may be of particular interest (CMB or other astrophysical emission), some may be unwanted (noise).

Obviously, components cannot be properly studied in data sets in which they appear only as a mixture. Component separation consists, for each of them, in isolating the emission from all the other components present in the data, in the best possible way.

It should be noted that what ``best" means depends on what the isolated data will be used for. Very often, one tries to obtain, for each component, an estimated map (or a set of maps at different frequencies) minimising the total error variance, i.e. minimising
\begin{equation}
\chi^2 = \sum_p |\widehat s(p) - s(p)|^2
\end{equation}
where $s(p)$ is the true component emission, and $\widehat s(p)$ its estimated value. $p$ indexes the space of interest for the component, typically a set of pixels $(\theta_p,\phi_p)$, or modes  $(\ell,m)$ of a spherical harmonic decomposition of a full sky map, or a set of Fourier modes $(k_x,k_y)$...

More generally, the objective of component separation is to estimate a \emph{set of parameters} which describes the component of interest. In the simplest case, this set of parameters may be emission in pixels, but it may also be instead parameters describing statistical properties such  as power spectra, spectral indices, etc\ldots  Since the set of parameters depends on the model assumed for the components, this model is of the utmost importance for efficient component separation. In the following, a significant part of the discussion will thus be dedicated to a summary of existing knowledge and of component modeling issues.

In the following, it is assumed that we are given a set of observations $y_i(p)$, where $i$, ranging from 1 to $N_\mathrm{chann}$, indexes the observation frequency. The observed emission in each of the frequency bands is supposed to result from a mixture of several astrophysical components, with additional instrumental noise.

In this review paper, we discuss in some detail the problem of diffuse component separation. 
The paper is organised as follows: in the next section, we review the principles and implementation of the ILC, a very simple method to average the measurements obtained at different frequencies; section \ref{sec:components} reviews the known properties of diffuse sky emissions, useful to model the observations and put priors on component parameters; section \ref{sec:cleaning} discusses observation and data reduction strategies to minimize the impact of foregrounds on CMB measurements, based on physical assumptions about the various emissions; section \ref{sec:linear-system} discusses the model of a linear mixture, and various options for its linear inversion to separate astrophysical components; section \ref{sec:maxent} discusses a non linear solution for inverting a linear mixture, based on a maximum entropy method; section \ref{sec:blind} presents the general ideas behind Blind Source Separation (BSS) and Independent Component Analysis (ICA); section \ref{sec:smica} discusses a particular method, the Spectral Matching ICA (SMICA); section \ref{sec:conclusion} concludes with a summary, hints about recent and future developments, open issues, and a prospective.

Let the reader be warned beforehand that this review may not do full justice to much of the work having been done on this very exciting topic. The discussion may be somewhat partial, although not intentionally. It has not been possible to the authors to review completely and compare all of the relevant work, partly for lack of time, and partly for lack of details in the published papers. As much as possible, we have nonetheless tried to mention all of the existing work, to comment the ideas behind the methods, and to quote most of the interesting and classical papers. 

\section{ILC: Internal Linear Combination} \label{sec:ilc}

The Internal Linear Combination (ILC) component separation method assumes very little about the components. One of them (e.g. the CMB) is considered to be the only emission of interest, all the other being unwanted foregrounds.

It is assumed that the template of emission of the component of interest is the same at all frequencies of observation, and that the observations are calibrated with respect to this component, so that for each frequency channel $i$ we have:
\begin{equation}
y_i(p) = s(p) + f_i(p) + n_i(p)
\label{eq:ILC-data-model}
\end{equation}
where $f_i(p)$ and $n_i(p)$ are foregrounds and noise contributions respectively in channel $i$.

A very natural idea, since all the observations actually measure $s(p)$ with some error $f_i(p) + n_i(p)$, consists in averaging all these measurements, giving a specific weight $w_i$ to each of them. 
Then, we look for a solution of the form:
\begin{equation}
\widehat s(p) = \sum_i w_i(p) y_i(p)
\end{equation}
where the weights $w_i(p)$ are chosen to maximize some criterion about the reconstructed estimate $\widehat s(p)$ of $s(p)$, while keeping the component of interest unchanged. This requires in particular that for all $p$, the sum of the coefficients $w_i(p)$ should be equal to 1.

\subsection{Simple ILC}

The simplest version of the ILC consists in minimising the variance $\sigma^2$ of the map $\widehat s(p)$ using weights independent of $p$ (so that $w_i(p) = w_i$ independent of $p$), with $\sum w_i = 1$. In this case, the estimated component is
\begin{eqnarray}
\widehat s(p) & = & \sum_i w_i y_i(p) \nonumber \\
& = & s(p) + \sum_i w_i f_i(p) + \sum_i w_i n_i(p)
.
\end{eqnarray}
Hence, under the assumption of de-correlation between $s(p)$ and all foregrounds, and between $s(p)$ and all noises, the variance of the error is minimum when the variance of the ILC map itself is minimum. 

\subsection{ILC implementation}\label{sub:ILC-implementation}

We now outline a practical implementation of the ILC method. For definiteness (and simplicity), we will assume here that the data is in the form of harmonic coefficients $s(\ell,m)$. 
The variance of the ILC map is:
\begin{equation}
\sigma^2 =  \sum_{\ell \geq 1}\sum_{m=-\ell}^\ell  \vec{w}\adj  \, {\m{C}}(\ell,m) \, {\vec{w}} 
=
\vec{w}\adj  \, {\m{C}} \, {\vec{w}}
\label{eq:sigma2-ILC}
\end{equation}
where ${\m{C}}(\ell,m) = \langle \vec{y}(\ell,m) \vec{y}^\dagger(\ell,m) \rangle$ is the covariance matrix of the observations in mode $(\ell,m)$, and ${\m{C}}$ is the covariance summed over all modes except $\ell=0$. $\vec{y}(\ell,m)$ and $\vec{w}$ stand for the vectors of generic element $y_i(\ell,m)$ and $w_i$ respectively.
The minimum, under the constraint of $\sum w_i =1$, is obtained, using the Lagrange multiplier method, by solving the linear system
\begin{eqnarray}
\forall i, \; \; \; \; \; \; \; \; \frac {\partial}{\partial w_i} \left [ \sigma ^2
+ \lambda \left( 1- \sum w_i \right) \right ]  & = & 0 \nonumber \\
\sum_i w_i & = & 1
\end{eqnarray}
Straightforward linear algebra gives the solution 
\begin{equation}
{w_i} = \frac{\sum_j \left [ {\m{C}}^{-1} \right ] _{ij}}{\sum_{ij} \left [ {\m{C}}^{-1} \right ] _{ij}}
\label{eq:ILC-weights}
\end{equation}
Note that if the template of emission of the component of interest is the same at all frequencies of observation, but the observations are {\emph{not}} calibrated with respect to this component, equation \ref{eq:ILC-data-model} becomes:
\begin{equation}
y_i(p) = a_is(p) + f_i(p) + n_i(p)
\label{eq:ILC-data-model2}
\end{equation}
In this case, it is still possible to implement an ILC. The solution is
\begin{equation}
\vec{w} = \frac{\vec{A}^T {\m{C}}^{-1} }{\vec{A}^T {\m{C}}^{-1} \vec{A}}
\label{eq:ILC-weights-general}
\end{equation}
where $\vec{A}$ is the vector of recalibration coefficients $a_i$. This solution of equation \ref{eq:ILC-weights-general} is equivalent to first changing the units in all the observations to make the response 1 in all channels, and then implementing the solution of equation \ref{eq:ILC-weights}.

\subsection{Examples of ILC separation: particular cases} \label{sub:ILC-examples}

This idea of ILC is quite natural. It has, however, several unpleasant features, which makes it non-optimal in most real-case situations. Before discussing this, let us examine now what happens in two simple particular cases.

\mypar{Case 1: Noisy observations with no foreground\\}

If there are no foregrounds, and the observations are simply noisy maps of $s(p)$, with independent noise for all channels of observation, the ILC solution should lead to a noise-weighted average of the measurements. 

Let us assume for simplicity that we have two noisy observations, $y_1$ and $y_2$, with $y_i=s+n_i$.
In the limit of very large maps, so that cross correlations between $s$, $n_1$ and $n_2$ vanish, the covariance matrix of the observations takes the form:
\[ 
\m{C} = 
\left[ 
\begin{array}{cc}
S+N_1 & S \\
S & S+N_2 \end{array} 
\right]\] 
where $S$ is the variance of the signal (map) of interest, and $N_1$ and $N_2$ the noise variances for the two channels. The inverse of $\m{C}$ is:
\[ 
{\mC}^{-1} = \frac{1}{\det (\m{C})}
\left[ 
\begin{array}{cc}
S+N_2 & -S \\
-S & S+N_1 \end{array} 
\right]\] 
and applying equation \ref{eq:ILC-weights}, we get $w_1=N_2/(N_1+N_2)$ and $w_2=N_1/(N_1+N_2)$. This is the same solution as weighting each map $i$ proportionally to $1/N_i$.

\mypar{Case 2: Noiseless observations with foregrounds\\}

Let us now examine the opposite extreme, where observations are noiseless linear mixtures of several astrophysical components. Consider the case of two components, with two observations. We can write the observations as $\vec{y} = \m{A}\vec{s}$, where $\m{A}$ is the so-called ``mixing matrix", and 
$\vec{s} = (s_1,s_2)^\dagger$ the vector of sources.

\noindent
The covariance of the observations is 
\[ {\m{C}} =  {\vec{y}} {\vec{y}} ^\dagger = {\mA} \, {\vec{s}}{\vec{s}}^\dagger {\mA}^\dagger\]
and its inverse is 
\begin{equation}
{\mC}^{-1} =  [{\mA}^\dagger]^{-1} \, [{\vec{s}}{\vec{s}}^\dagger]^{-1} {\mA}^{-1}
\label{eq:ILC-ex2-invcov}
\end{equation}
Let us assume that we are interested in the first source. The data are then calibrated so that the mixing matrix $\mA$ and its inverse are of the form
$$
\mA = \left[ 
\begin{array}{cc}
1 & a_{12}\\
1 & a_{22} \end{array} 
\right] 
\; \; \; \; \; \; \rm{and} \; \; \; \; \; \;
{\mA}^{-1} = \frac{1}{\det ({\mA})} \left[ 
\begin{array}{cc}
a_{22} & -a_{12}\\
-1 & 1 \end{array} 
\right] 
$$
Then, if we assume that components 1 and 2 are uncorrelated, equation \ref{eq:ILC-ex2-invcov} yields
\begin{equation}
{\mC}^{-1}  = \frac{1}{(\det ({\mA}))^2}
\left[ 
\begin{array}{cc}
a_{22} & -1 \\
-a_{12} & 1
\end{array}
\right] 
\left[ 
\begin{array}{cc}
S_1^{-1} & 0 \\
0 & S_2^{-1}
\end{array}
\right] 
\left[ 
\begin{array}{cc}
a_{22} & -a_{12}\\
-1 & 1 \end{array} 
\right]
\label{eq:C-1form} 
\end{equation}
where $S_1$ and $S_2$ are the variances of components 1 and 2 respectively. After expansion of the matrix product, we get:
\begin{equation}
{\mC}^{-1}  = \frac{1}{(\det ({\mA}))^2}
\left[ 
\begin{array}{cc}
(a_{22}^2S_1^{-1} + S_2^{-1}) & (-a_{22}a_{12}S_1^{-1} - S_2^{-1})\\
(-a_{22}a_{12}S_1^{-1} - S_2^{-1}) & (a_{12}^2S_1^{-1} + S_2^{-1}) \end{array} 
\right] 
\end{equation}
and using equation \ref{eq:ILC-weights}, we get
\begin{equation}
{\vec{w}} = \frac{1}{(a_{22}-a_{12})} \left[ 
\begin{array}{c}
a_{22}\\
-a_{12} \end{array} 
\right] 
\end{equation}
which is the transpose of the first line of matrix ${\mA}^{-1}$, so that $\widehat s_1 = {\vec{w}}.{\vec{y}} = s_1$. As expected, \emph{if the covariance of the two components vanishes}, the ILC solution is equivalent, for the component of interest, to what is obtained by inversion of the mixing matrix.

What happens now if the two components are correlated? Instead of the diagonal form ${\rm{diag}}(S_1,S_2)$, the covariance matrix of the sources contains an off-diagonal term $S_{12}$, so that equation \ref{eq:C-1form} becomes:
\begin{equation}
{\mC}^{-1}  = \frac{1}{(\det ({\mA}))^2} \frac{1}{(S_1S_2 - S_{12}^2)}
\left[ 
\begin{array}{cc}
a_{22} & -1 \\
-a_{12} & 1
\end{array}
\right] 
\left[ 
\begin{array}{cc}
S_2 & S_{12} \\
S_{12} & S_1
\end{array}
\right] 
\left[ 
\begin{array}{cc}
a_{22} & -a_{12}\\
-1 & 1 \end{array} 
\right]
\label{eq:C-1form-correl} 
\end{equation}
which yields the solution
\begin{equation}
{\vec{w}} = \frac{1}{(a_{22}-a_{12})} \left[ 
\begin{array}{c}
a_{22} + S_{12}/S_2\\
-a_{12} - S_{12}/S_2 \end{array} 
\right] 
\end{equation}
The ILC is not equivalent anymore to the inversion of the mixing matrix $\mA$. Instead, the estimate $\widehat s_1$
of $s_1$ is:
\begin{equation}
\widehat s_1 = {\vec{w}}.{\vec{y}} = s_1 - \frac{S_{12}}{S_2} s_2
\end{equation}
The ILC result is biased, giving a solution in which a fraction of $s_2$ is subtracted erroneously, in proportion to the correlation between $s_1$ and $s_2$, to lower as much as possible the variance of the output map. The implication of this is discussed in paragraph \ref{sub:comments-ILC}.

\subsection{Improving the ILC method}\label{sub:improvements-ILC}

With the exception of the CMB, diffuse sky emissions are known to be very non stationnary (e.g. galactic foregrounds are strongly concentrated in the galactic plane). In addition, most of the power is concentrated on large scales (the emissions are strongly correlated spatially).
As the ILC method minimizes the total variance of the ILC map (the integrated power from all scales, as can be seen in equation \ref{eq:sigma2-ILC}), the weights $w_i$ are strongly constrained essentially by regions of the sky close to the galactic plane, where the emission is strong, and by large scales, which contain most of the power. In addition, the ILC method finds weights resulting from a compromise between reducing astrophysical foreground contamination, and reducing the noise contribution. In other words, for a smaller variance of the output map, it pays off more to reduce the galactic contamination in the galactic plane and on large scales, where it is strong, rather than at high galactic latitude and on small scales, where there is little power anyway. This particularity of the ILC when implemented globally is quite annoying for CMB studies, for which all scales are interesting, and essentially the high galactic latitude data is useful.

Away from the galactic plane and on small scales, the best linar combination for cleaning the CMB from foregrounds and noise may be very different from what it is close to the galactic plane and on large scales. A very natural idea to improve on the ILC is to decompose sky maps in several regions and/or scales, and apply an ILC independently to all these maps. The final map is obtained by adding-up all the ILC maps obtained independently in various regions and at different scales. Applications of these ideas are discussed in the next paragraph.

\subsection{ILC-based foreground-cleaned CMB map from WMAP data}

A map of CMB anisotropies has been obtained using the ILC method \cite{2003ApJS..148...97B} on first year data from the WMAP mission, and has been released to the scientific community as part of the first year WMAP data products.

The input data is the set of five all sky, band averaged maps for the K, Ka, Q, V and W frequency bands, all of which smoothed to the same 1 degree resolution for convenience. The ILC is performed independently in 12 regions, 11 of which being in the WMAP kp2 mask at low galactic latitudes, designed to mask out regions of the sky highly contaminated by galactic foregrounds. This division into twelve regions is justified by the poor performance of the ILC on the full sky, interpreted as due to varying spectral indices of the astrophysical foregrounds. Discontinuities between the regions are reduced by using smooth transitions between the regions.

Little detail is provided on the actual implementation of the ILC by the WMAP team. Apparently, a non-linear iterative minimization algorithm was used, instead of the linear solution outlined in paragraph \ref{sub:ILC-implementation}. Although there does not seem to be any particular reason for this choice, in principle the particular method chosen to minimize the variance does not matter, as long as it finds the minimum efficiently. There seem to be, however, indications that the convergence was not perfect, as discussed by Eriksen and collaborators in a paper discussing the ILC and comparing the results of the several of its implementations on WMAP data \cite{2004ApJ...612..633E}. Caution should probably be taken when using the WMAP ILC map for any purpose other than a visual impression of the CMB.

Tegmark and collaborators have improved the ILC method in several respects, and provide an independent CMB map obtained from WMAP data by ILC \cite{2003PhRvD..68l3523T}. Their implementation allows the weights to depend not only on the region of the sky, but also on angular scale, as discussed in paragraph \ref{sub:improvements-ILC}. In addition, they partially deconvolve the WMAP maps in harmonic space to put them all to the angular resolution of the channel with the smallest beam, rather than smoothing all maps to put them all to the angular resolution of the channel with the largest beam. As a result, their ILC map has better angular resolution, but higher total noise. The high resolution map, however, can be filtered using a Wiener filter for minimal variance of the error. The Wiener-filtered map is obtained by multiplying each $a_{\ell m}$ mode of the map by a factor 
$$W(\ell,m) = C_\ell/S_{\ell}$$
where $C_\ell$ is the estimated CMB power spectrum (computed for the cosmological model fitting best the WMAP data estimate), and $S_{\ell}$ is the estimated power spectrum of the noisy CMB map obtained by the authors using their ILC method.

\begin{figure} 
\centering
\includegraphics[width=0.8\textwidth]{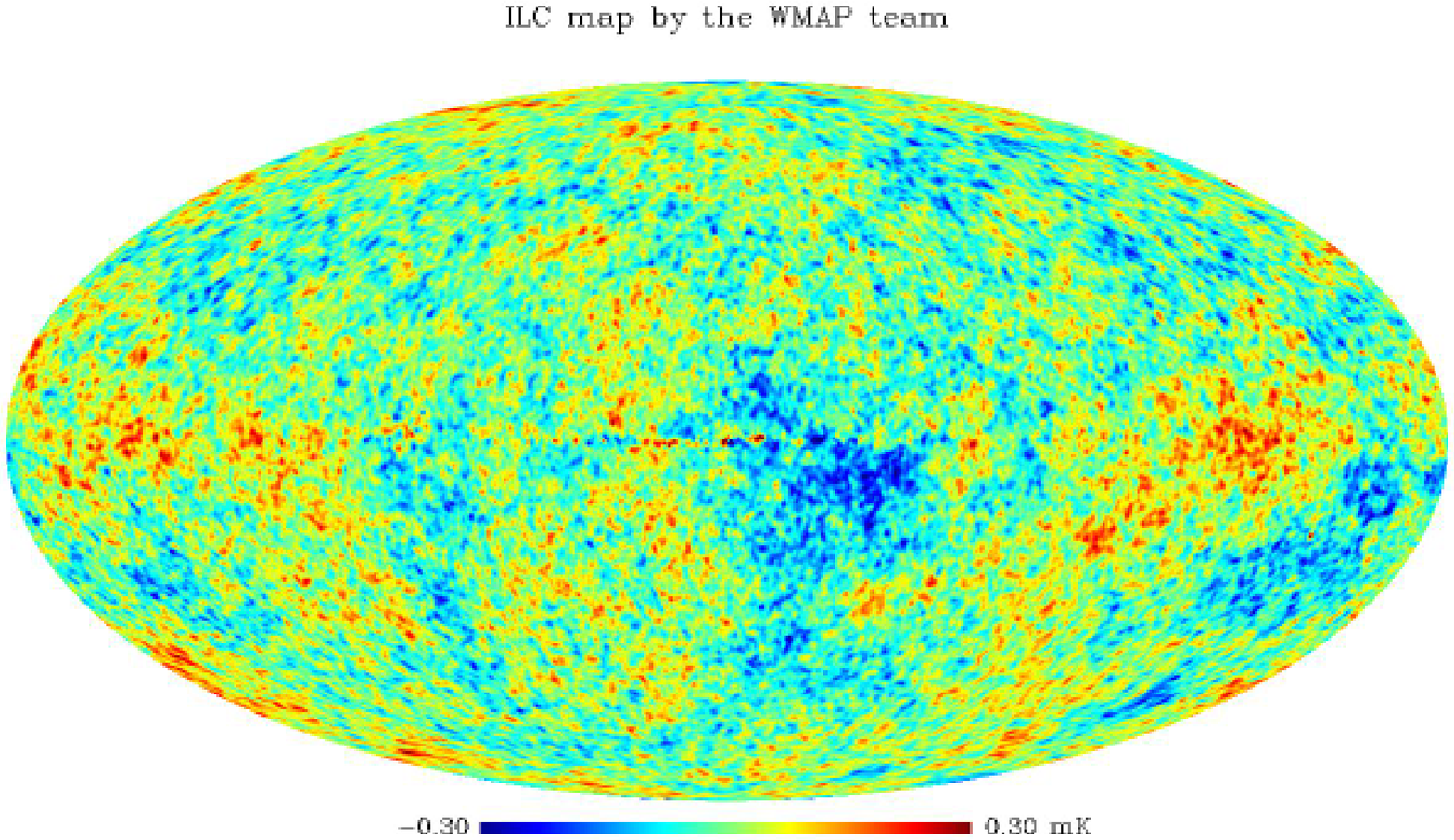} 
\caption{The ILC map of the CMB obtained by the WMAP team (one year data). Residuals of galactic emission are clearly visible in the center of the map. The color scale spans a range of -300 to +300 $\mu$K thermodynamic, although localised residuals exceed these values.} 
\label{fig:ILC-WMAP}
\centering
\includegraphics[width=0.8\textwidth]{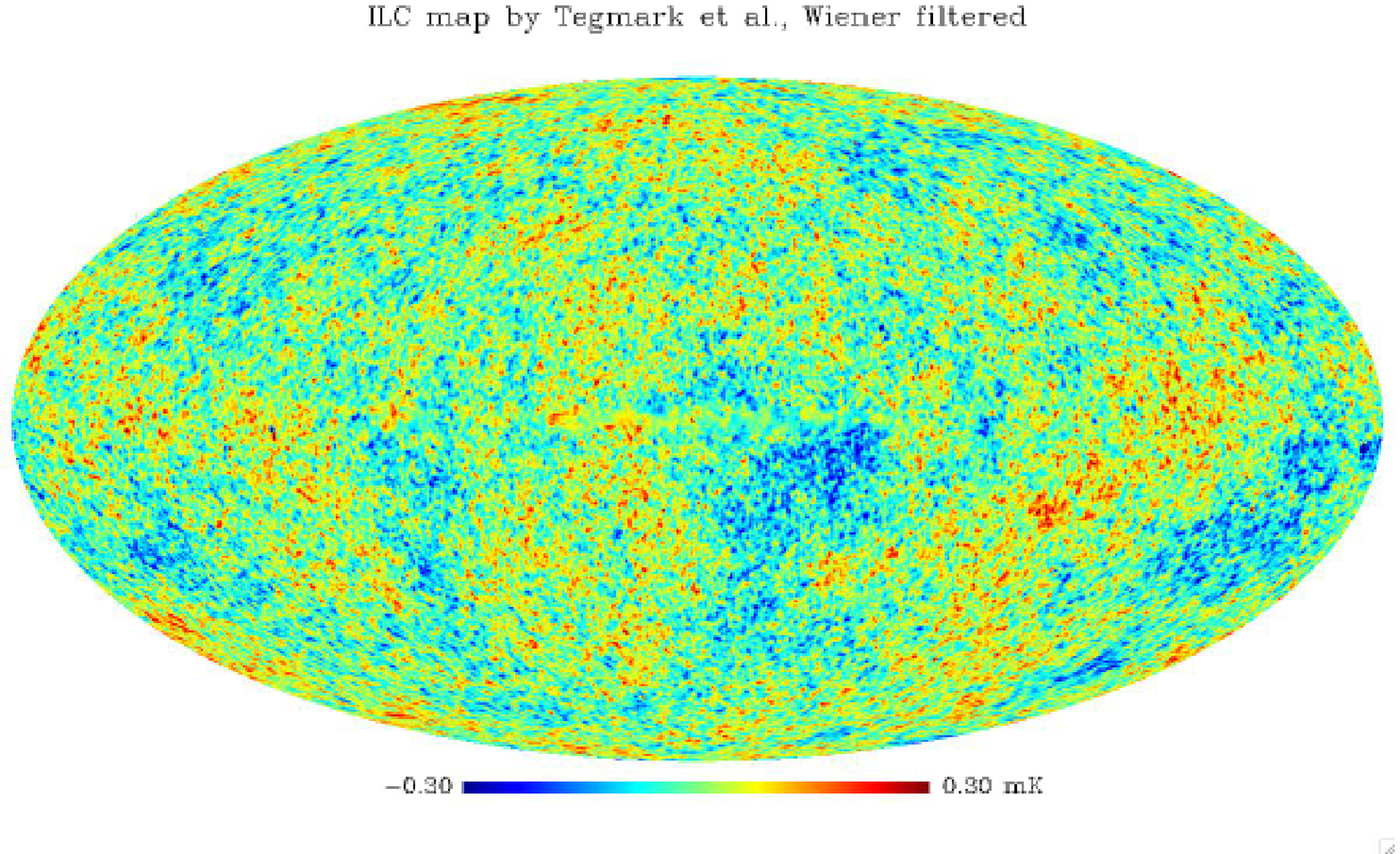} 
\caption{The foreground-cleaned CMB map of Tegmark et al., obtained by the ILC method described in \cite{2003PhRvD..68l3523T}, after Wiener filtering. The effect of the region and scale-dependent weighting can be seen in the center of the map (galactic center) where the map looks smoother and flatter than elsewhere. The color scale spans a range of -300 to +300 $\mu$K, although localised residuals exceed these values, as in figure \ref{fig:ILC-WMAP}. The superior angular resolution can clearly be seen.} 
\label{fig:ILC-Tegmark}
\end{figure}

The CMB map obtained by the WMAP team from first year data is shown in figure \ref{fig:ILC-WMAP}. For comparison, the map obtained by Tegmark et al. is shown in figure \ref{fig:ILC-Tegmark}. Both give a good visual perception of what the CMB field looks like. 

\subsection{Comments about the ILC}\label{sub:comments-ILC}

The ILC has been used essentially to obtain a clean map of CMB emission. In principle, nothing prevents using it also for obtaining cleaned maps of other emissions, with the caveats that 
\begin{itemize}
\item The data must be calibrated with respect to the emission of interest, so that the data takes the form of equation \ref{eq:ILC-data-model}. This implies that the template of emission of the component of interest should not change significantly with the frequency-band of observation. This is the case for the CMB (temperature and polarisation), or for the SZ effect (to first order at least... more on this later).
\item The component of interest should not be correlated with other components. Galactic components, being all strongly concentrated in the galactic plane, can thus not be recovered reliably with the ILC.
\end{itemize}
This issue of decorrelation of the component of interest $s(p)$ and the foregrounds can also generate problems in cases where the empirical correlation between the components does not vanish.  As demonstrated in \ref{sub:ILC-examples}, the ILC method will not work properly (biasing the result) if the assumption the component of interest $s(p)$ are correlated for whatever reason.
In particular, small data sets, even if they are realisations of actually uncorrelated random processes, are always empirically correlated to some level. For this reason, the ILC should not be implemented independently on too small subsets of the original data (very small regions, very few modes).

Finally, whereas the ILC is a powerful tool  when nothing is known about the data, it is certainly non optimal when prior information is available. Foreground emissions are discussed in some detail in following section.

\section{Sky emission model: components} \label{sec:components}

``Know your enemy"... This statement, borrowed from elementary military wisdom, applies equally well in the fight against foreground contamination. Prior knowledge about astrophysical components indeed has been widely used in all practical CMB data analyses. Methods can then be specifically tailored to remove foregrounds based on their physical properties, in particular their morphology, their localisation, and their frequency scaling based on the physical understanding of their emission mechanisms.

In addition to knowledge about the unwanted foregrounds, prior knowledge about the component of interest is of the utmost importance for its identification and separation in observations. In the ILC method discussed above, for instance, the prior knowledge of the emission law of the CMB (derivative of a blackbody) is specifically used.

\subsection{The various astrophysical emissions}

Astrophysical emissions relevant to the framework of CMB observations can be classified in three large categories (in addition to the CMB itself). Diffuse galactic emission, extragalactic emission, and solar system emission.

Diffuse galactic emissions originate from the local interstellar medium (ISM) in our own galaxy. The ISM is constituted of cold clouds of molecular or atomic gas, of an intercloud medium which can be partly ionised, and of hot ionized regions presumably formed by supernovae. These different media are strongly concentrated in the galactic plane.
The intensity of corresponding emissions decreases with galactic latitude with a cosecant law behaviour (the optical depth of the emitting material scales proportionnally to $1/\sin b$).
Energetic free electrons spiralling in the galactic magnetic field generate synchrotron emission, which is the major foreground at low frequencies (below a few tens of GHz). Warm ionised material emits free-free (Bremstrahlung) emission, due to the interaction of free electrons with positively charged nuclei. Small particles of matter (dust grains and macromolecules) emit radiation as well, through thermal greybody emission, and possibly through other mechanisms.

Extragalactic emissions arise from a large background of resolved and unresolved radio and infrared galaxies, as well as clusters of galaxies. The thermal and kinetic Sunyaev-Zel'dovich effects, due to the inverse Compton scattering of CMB photons off hot electron gas in ionized media, are of special interest for cosmology. These effects occur, in particular, towards clusters of galaxies, which are known to comprise a hot (few keV) electron gas. Infrared and radiogalaxies emit also significant radiation in the frequency domain of interest for CMB observations, and contribute both point source emission from nearby bright objects, and a diffuse background due to the integrated emission of a large number of unresolved sources, too faint to be detected individually, but which contribute sky background inhomogeneities which may pollute CMB observations.

Solar system emission comprises emissions from the planets, their satellites, and a large number of small objects (asteroids). In addition to those, there is diffuse emission due to dust particles and grains in the ecliptic plane (zodiacal light). The latter is significant essentially at the highest frequencies of an instrument like the Planck HFI \cite{2006A&A...452..685M}.

In the rest of this section, we briefly outline the general properties of these components and the modeling of their emission in the centimetre to sub-millimetre wavelength range.

\subsection{The Cosmic Microwave Background}
The cosmic microwave background, relic radiation from the hot big bang emitted at the time of decoupling when the Universe was about 370,000 years old, is usually thought of (by cosmologists) as the component of interest in the sky emission mixture. Millimetre and submillimetre wave observations, however, sometimes aim not only at measuring CMB anisotropies, but also other emissions. In this case, the CMB becomes a noxious background which has to be subtracted out of the observations, just as any other.

The CMB emission is relatively well known already. The main theoretical framework of CMB emission can be found in any modern textbook on cosmology, as well as in several reviews \cite{2002ARA&A..40..171H,2002AmJPh..70..106W}. The achievement of local thermal equilibrium in the primordial plasma before decoupling, together with the very low level of the perturbations, guaranties that CMB anisotropies are properly described as the product of a spatial template $\Delta T(p) = T_{\rm CMB}(p) - \overline T_{\rm CMB}$, and a function of $\nu$ (frequency scaling) which is the derivative of a blackbody with respect to temperature:
\begin{equation}
\Delta I_\nu(p) = \Delta T_{\rm CMB}(p) \left [  \frac{\partial B_\nu(T)}{\partial T} \right ] _{T = \overline T_{\rm CMB} \simeq 2.726 \, {\rm{K}}}
\end{equation}
In the standard cosmological model, the CMB temperature fluctuation map $\Delta T(p)$ is expected to be a realisation of a stationary Gaussian random field, with a power spectrum $C_\ell$ displaying a series of peaks and troughs (the acoustic peaks), the location and relative size of which are determined by a few free parameters of the cosmological model.\footnote{The power spectrum $C_\ell$ is defined as the set of variances of the coefficients $a_{\ell m}$ of the expansion of the random field representing CMB relative temperature fluctuations $\Delta T(p) / \overline T_{\rm CMB}$ onto the basis of spherical harmonics on the sphere $Y_{\ell m}(\theta,\phi)$. The stationarity and isotropy of the random field guarantees that the variance of $a_{\ell m}$ (coefficients $C_\ell$) is independent of $m$.}

Good maps of sky emission at a resolution of about 15 arcminutes, obtained from WMAP data in the frequency range 20--90 GHz, clearly comprise at high galactic latitude an astrophysical component compatible with all these predictions. The power spectrum is measured with excellent accuracy by WMAP up to the second Doppler peak, while complementary balloon--borne and ground--based experiments yield additional measurements at higher $\ell$ (smaller scales).

Efficient diffuse component separation methods should make use of this current status of knowledge about the CMB:
\begin{itemize}
\item Law of emission, known to a high level of precision to be the derivative of a blackbody with respect to temperature, as expected theoretically and checked experimentally with the Boomerang \cite{2000Natur.404..955D} and Archeops \cite{2005A&A...436..785T} multifrequency data sets, as well as with the WMAP data \cite{2003ApJS..148....1B,2005MNRAS.364.1185P} 
\item Stationnarity and gaussianity to a high level of accuracy, as expected theoretically and checked on WMAP data \cite{2003ApJS..148..119K} 
\item Good cosmological prior on the power spectrum of the fluctuations, validated experimentally with several data sets \cite{2002ApJ...571..604N,2006astro.ph..3451H}
\end{itemize}

\begin{figure} [thb]
\centering
\includegraphics[width=\textwidth]{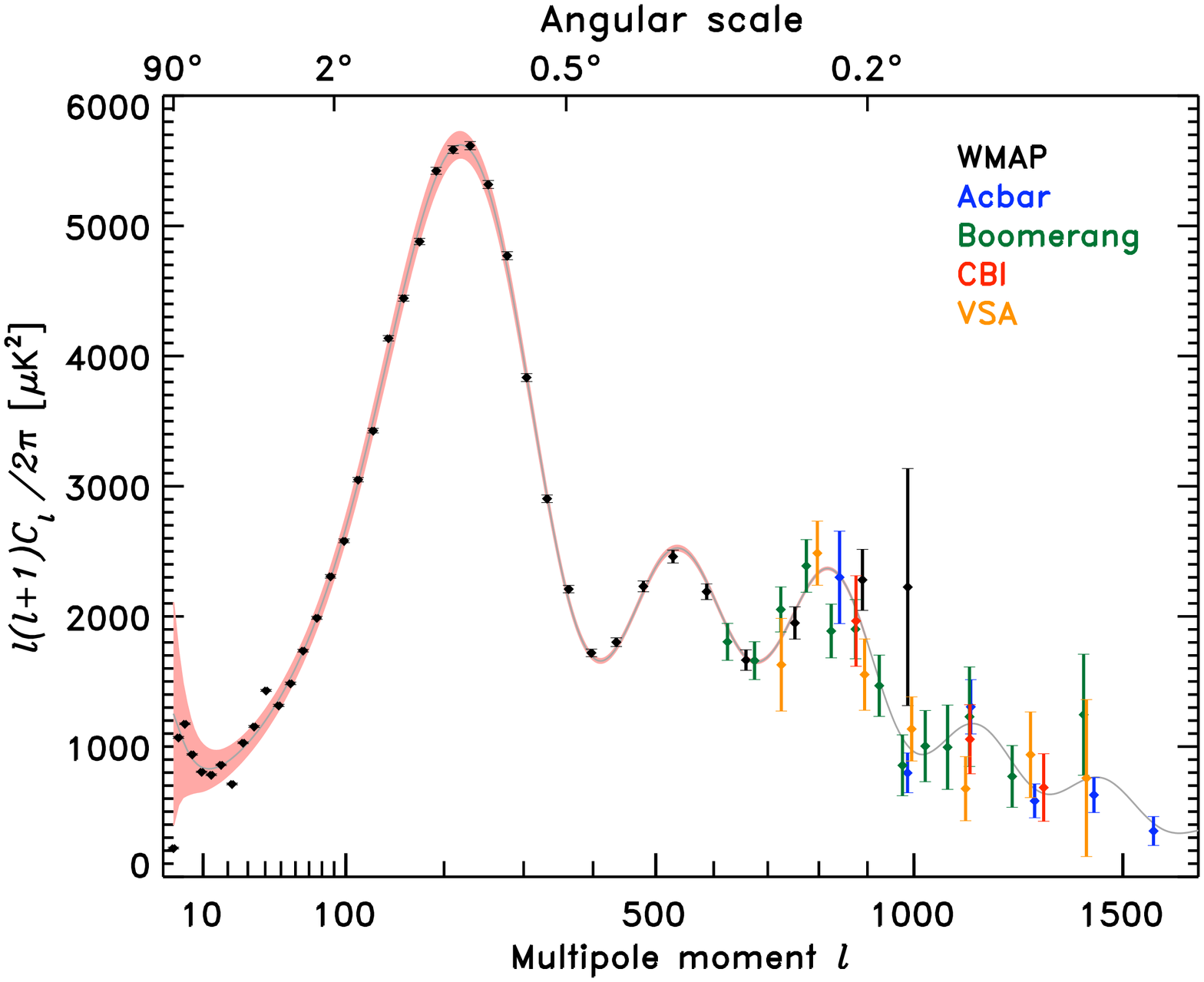} 
\caption{Present-day best constraints of the CMB temperature power spectrum (from \cite{2006astro.ph..3451H}). Data sets in addition to WMAP 3--year data are from \cite{2005astro.ph..7494J,2004ApJ...600...32K,2004ApJ...609..498R,
2004MNRAS.353..732D}} 
\label{fig:WMAP-3yr-CMB-spec}
\end{figure}

A good visual impression of all-sky CMB emission is given in figures \ref{fig:ILC-WMAP} and \ref{fig:ILC-Tegmark}. The present status of knowledge of the power spectrum $C_\ell$ is shown in figure \ref{fig:WMAP-3yr-CMB-spec}.

The extraction of CMB emission from a set of multifrequency observations may be done with the following objectives in mind (at least):
\begin{itemize}
\item Get the best possible map of the CMB (in terms of total least square error, from noise and foregrounds together);
\item Get the CMB map with the least possible foreground contamination;
\item Get the CMB map for which spurious non-gaussianity from foregrounds, noise and systematic effects is minimal;
\item Get the best possible estimate of the CMB angular power spectrum...
\end{itemize}
Obviously, the best component separation method for extracting the CMB will depend on which of the above is the primary objective of component separation.

\subsection{Emissions from the diffuse interstellar medium}

\mypar{Synchrotron emission\\}

Synchrotron emission arises from energetic charged particles spiralling in a magnetic field. In our galaxy, such magnetic fields extend outside the galactic plane. Energetic electrons originating from supernovae shocks, spiraling in this field, can depart the galactic plane and generate emission even at high galactic latitudes. For this reason, synchrotron emission is less concentrated in the galactic plane than free-free and dust.

The frequency scaling of synchrotron emission is a function of the distribution of the energies of the radiating electrons. For number density distributions $N(E) \propto E^{-\gamma}$, the flux emission is also in the form of a power law, $I_\nu \propto \nu^{-\alpha}$, with $\alpha = (\gamma -1)/2$.
In Rayleigh-Jeans (RJ) temperature $\Delta T \propto  \nu^{-\beta}$ with $\beta = \alpha+2$. Typically, $\beta$ ranges from 2.5 to 3.1, and is somewhat variable across the sky.

In spite of a moderate sensitivity for current standards, the 408 MHz all sky map \cite{1981A&A...100..209H}, dominated by synchrotron emission, gives a good visual impression of the distribution of synchrotron over the sky.

\begin{figure} [thb]
\centering
\includegraphics[width=\textwidth]{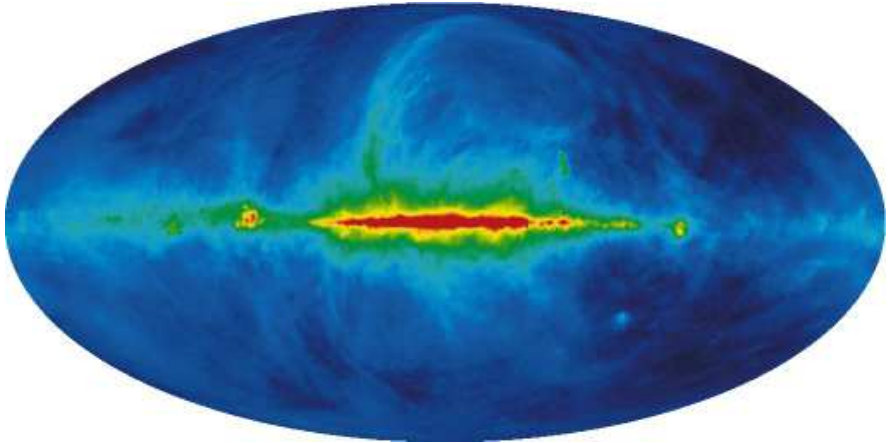} 
\caption{The 408 MHz all-sky synchrotron map \cite{1981A&A...100..209H}. Data and images are available on the NASA Lambda web site.}
\label{fig:haslam}
\end{figure} 

In principle, synchrotron emission can be highly polarised, up to 50-70\%

\mypar{Free-Free emission\\}

Free-free emission is the least well known observationally of the three major emissions originating from the galactic interstellar medium in the millimetre and centimetre wavelength range. This emission arises from the interaction of free electrons with ions in ionised media, and is called ``free-free" because of the unbound state of the incoming and outgoing electron. Alternatively, free-free is called ``Bremsstrahlung" emission (``braking radiation" in German), because photons are emitted while electrons loose energy by interaction with the heavy ions.

Theoretical calculations of free-free emission in an electrically neutral medium consisting of ions and electrons gives an estimate of the brightness temperature at frequency $\nu$ for free-free emission of the form:
\begin{equation}
T_{\rm ff} \simeq 0.08235 \; T_e^{-0.35} \, \nu^{-\beta} \int_{\rm l.o.s.} N_e N_i dl
\end{equation}
where $T_e$ is in Kelvin, $\nu$ is in GHz and the line of sight integral of  electron and ion density in cm$^{-6}$pc \cite{1998astro.ph..1121S}. Theoretical estimates of the spectral index, $\beta$, range from about $2.1$ to $2.15$, with errors of $\pm 0.03$.

While free-free emission is not observed directly, as it never dominates over other astrophysical emissions, the source of its emission (mainly ionised hydrogen clouds) can be traced with hydrogen recombination emission lines, and particularly H$\alpha$ emission. The connection between H$\alpha$ and free-free has been discussed extensively by a number of authors \cite{1998astro.ph..1121S,1998PASA...15..111V,1999ASPC..181..253M}. We have:

\begin{equation}
\frac{T_{\rm ff}[{\rm mK}]}{I_\alpha [{\rm R}]} \simeq 10.4 \, \nu^{-2.14} \, T_4^{0.527} \, 10^{0.029/T_4} \,
(1+0.08)
\end{equation}
Where $T_{\rm ff}[{\rm mK}]$ is the free-free brightness temperature in mK, $I_\alpha [{\rm R}]$ the H$\alpha$ surface brightness in Rayleigh, $\nu$ the frequency, and $T_4$ the temperature of the ionized medium in units of $10^4$ K. The Rayleigh (R) is defined as 1 R = $(10^6/4\pi)$ photons/cm$^2$/s/sr.

Free-free emission, being due to incoherent emissions from individual electrons scattered by nuclei in a partially ionised medium, is not polarised (to first order at least).

\mypar{Thermal emission of galactic dust\\}
The present knowledge of interstellar dust is based on extinction observations from the near infrared to the UV domain, and on observations of its emission from radio frequencies to the infrared domain.

Dust consists in small particles of various materials, essentially silicate and carbonaceous grains of various sizes and shapes, in amorphous or crystalline form, sometimes in aggregates or composites. Dust is thought to comprise also large molecules of polycyclic aromatic hydrocarbon (PAH). The sizes of the grains range from few nanometers for the smallest, to micrometers for the largest. They can emit through a variety of mechanisms. The most important for CMB observations is greybody emission in the far infrared, at wavelengths ranging from few hundreds of microns to few millimeters. The greybody emission is typically characterised by a temperature $T_{\rm dust}$ and by an emissivity proportional to a power of the frequency $\nu$:
\begin{equation}
I_\nu \propto \nu^\beta B_\nu(T_{\rm dust})
\end{equation}
where $B_\nu(T)$ is the usual blackbody emission
\begin{equation}
B_\nu(T) = \frac{2h\nu^3}{c^2} \frac{1}{e^{h\nu/kT}-1}
\end{equation}
This law is essentially empirical. In practice, dust clouds along the line of sight can have different temperatures and different compositions: bigger or smaller grains, different materials. They can thus have different emissivities as well.
Temperatures for interstellar dust are expected to range from about 5 Kelvin to more than 30 Kelvin, depending on the heating of the medium by radiation from nearby stars, with typical values of 16-18 K for emissivity indices $\beta \simeq 2$.

In principle, thermal emission from galactic dust should not be strongly polarised, unless dust particles are significantly asymmetric (oblate or prolate), and there exists an efficient process for aligning the dust grains in order to create a significant statistical asymmetry. Preliminary dust observations with the Archeops instrument \cite{2004A&A...424..571B, 
2005A&A...444..327P} seems to indicate polarisation levels of the order of few per cent, and as high as 15-20 per cent in a few specific regions.

\mypar{Spinning dust or anomalous dust emission}

In the last years, increasing evidence for dust-correlated emissions at frequencies below 30 GHz, in excess to expectations from synchrotron and free-free, has aroused interest in possible non-thermal emissions from galactic dust \cite{1996ApJ...460....1K,1997ApJ...486L..23L}. Among the possible non-thermal emission mechanisms, spinning dust grains offer an interesting option \cite{1998ApJ...494L..19D}.

At present, there still is some controversy on whether the evidence for non-thermal dust emission is robust enough for an unambiguous statement. Observations of different sky regions, indeed, yield somewhat different results \cite{2006ApJ...643L.111D,2006MNRAS.370...15F}, which may be due either to varying dust cloud properties, or to differences in the analyses and interpretations, or both. Certainly, investigating this question is an objective of primary interest for diffuse component separation methods (especially blind ones) in the near future.

\subsection{The SZ effect}

The Sunyaev Zel'dovich (SZ) effect \cite{1972CoASP...4..173S} is the inverse Compton scattering of CMB photons on free electrons in ionised media. In this process, the electron gives a fraction of its energy to the scattered CMB photon.
There are, in fact, several SZ effects: The thermal SZ effect is due to the scattering of photons on a high temperature electron gas, such as can be found in clusters of galaxies. The kinetic SZ effect is due to the scattering on a number of electrons with a global radial bulk motion with respect to the cosmic background. Finally, the polarised SZ effect is a second order effect due to the kinematic quadrupole of the CMB in the frame of an ensemble of electrons with a global transverse bulk motion with respect to the CMB.

SZ effects are not necessarily linked to clusters of galaxies. Any large body with hot ionised gas can generate significant effects. It has been proposed that signatures of inhomogeneous reionisation can be found via the kinetic and thermal SZ effect \cite{1996A&A...311....1A,1998ApJ...508..435G,1999ApJ...522...66Y}. However, the largest expected SZ signatures originate from ionised intra-cluster medium.

\mypar{Clusters of galaxies\\}
Clusters of galaxies, the largest known massive structures in the Universe, form by gravitational collapse of matter density inhomogeneities on large scales (comoving scales of few Mpc). They can be detected either optically from concentrations of galaxies at the same redshift, or in the submillimeter by their thermal SZ emission, or by the effect of their gravitational mass in weak shear maps, or in X-ray. The hot intracluster baryonic gas can be observed through its X--ray emission due to Bremsstrahlung (free-free) emission of the electrons on the nuclei, which permits to measure the electron temperature (typically a few keV). On the sky, typical cluster angular sizes range from about one arcminute to about one degree. Clusters are scattered over the whole sky, although this distribution follows the repartition of structure on the largest scales in the universe. Large scale SZ effect observations may be also used to survey the distribution of hot gas on these very large scales, although such SZ emission, from filaments and pancakes in the distribution, is expected to be at least an order of magnitude lower in intensity than thermal SZ emission from the clusters themselves.

Each cluster of galaxies has its own thermal, kinetic and polarised SZ emission. These various emissions and their impact on CMB observations and for cosmology have been studied by a variety of authors. Useful reviews have been made by Birkinshaw \cite{1999PhR...310...97B} and Rephaeli \cite{2002AIPC..616..309R}, for instance.

\mypar{Thermal SZ\\}

The thermal SZ effect generated by a gas of electrons at temperature $T_e$ is, in fact, a spectral distortion of the CMB emission law. It is common to consider as the effect the difference 
$\Delta I_\nu = I_\nu - B_\nu (T_{\rm CMB})$ 
between the distorted CMB photon distribution $I_\nu$ and the original one $B_\nu (T_{\rm CMB})$. 
In the non-relativistic limit (when $T_e$ is lower than about 5 keV, which is the case for most clusters), the shape of the spectral distortion does not depend on the temperature. The change in intensity due to the thermal SZ effect is:
\begin{equation}
\Delta I_\nu = y  \frac{x e^x}{(e^x - 1)} \left [  \frac{x(e^x + 1)}{(e^x - 1)} - 4 \right ] B_\nu(T_{\rm CMB})
\end{equation}
where $B_\nu(T_{\rm CMB})$ is the Planck blackbody emission law at CMB temperature
$$
B_\nu(T_{\rm CMB}) = \frac{2h\nu^3}{c^2} \frac{1}{e^x -1}
$$
and $x=h\nu/kT_{\rm CMB}$. The dimensionless parameter $y$ (Comptonisation parameter) is proportional to the integral of the electron pressure along the line of sight:
$$
y = \int_{\rm los} \frac{kT_e}{m_ec^2} n_e \sigma_{\rm thomson} dl
$$
where $T_e$ is the electron temperature, $m_e$ the electron mass, $c$ the speed of light, $n_e$ the electron density, and $\sigma_{\rm thomson}$ the Thomson cross section.

\mypar{Kinetic SZ\\}

The kinetic SZ effect is generated by the scattering of CMB photons off an electron gas in motion with respect to the CMB. This motion generates spectral distortions with the same frequency scaling as CMB temperature fluctuations, and are directly proportionnal to the velocity of the electrons along the line of sight. As the effect has the same frequency scaling as CMB temperature fluctuations, it is, in principle, indistiguishable from primordial CMB. However, since the effect arises in clusters of galaxies with typical sizes 1 arcminute, it can be distinguished to some level from the primordial CMB by spatial filtering, especially if the location of the clusters most likely to generate the effect is known from other information (e.g. the detection of the clusters through the thermal SZ effect).



\mypar{Polarised SZ\\}

The polarised SZ effect arises from the polarisation dependence of the Thomson cross section:
$$
\sigma_T \propto \left | \vec{e_1}.\vec{e_2} \right |^2
$$
where $\vec{e_1}$ and $\vec{e_2}$ are the polarisation states of the incoming and outgoing photon respectively.
A quadrupole moment in the CMB radiation illuminating the cluster electron gas generates a net polarisation, at a level typically two orders of magnitude lower than the kinetic SZ effect \cite{1980MNRAS.190..413S,1999MNRAS.305L..27A,1999MNRAS.310..765S}. Therefore, the kinetic SZ effect has been proposed as a probe to investigate the dependence of the CMB quadrupole with position in space. Cluster transverse motions at relativistic speed, however, generate also such an effect from the kinematic quadrupole induced by the motion. Multiple scattering of CMB photons also generates a low-level polarisation signal towards clusters.

The polarised SZ effects has a distinctive frequency scaling, independent (to first order) to cluster parameters and to the amplitude of the effect. Amplitudes are proportionnal:
\begin{itemize}
\item to $\tau$  for the intrinsic CMB quadrupole effect,
\item to $(v_t/c)^2 \tau$ for the kinematic quadrupole effect
\item to $(kT_e/m_ec^2)\tau^2$ and $(v_t/c)\tau^2$ for polarisation effects due to double scattering. 
\end{itemize}
Here $\tau$ is the optical depth, $v_t$ the transverse velocity, $c$ the speed of light, $k$ the Boltzmann constant, and $T_e$ and $m_e$ the electron temperature and mass.

As polarised effects arise essentially in galaxy clusters, they can be sought essentially in places where the much stronger thermal effect is detected, which will permit to improve the detection capability significantly. Polarised SZ emission, however, is weak enough that it is not expected to impact significantly the observation of any of the main polarised emissions.

\mypar{Diffuse component or point source methods for SZ effect separation?\\}
The SZ effect is particular in several respects. As most of the emission comes from compact regions towards clusters of galaxies (at arcminute scales), most of the present-day CMB experiments do not resolve clusters individually (apart for a few known extended clusters). For this reason, it seems natural to use point source detection methods for cluster detection (see review by Barreiro \cite{barreiro-compact-sources}). However, the very specific spectral signature, the presence of a possibly large background of clusters with emission too weak for individual cluster detection, and the interesting possibility to detect larger scale diffuse SZ emission, makes looking for SZ effect with diffuse component separation methods an interesting option.

\subsection{The infrared background of unresolved sources}

The added emissions from numerous unresolved infrared sources at high redshift make a diffuse infrared background, detected originally in the FIRAS and DIRBE data \cite{1996A&A...308L...5P}. Because each source has its specific emission law, and because this emission law is redshifted by the cosmological expansion, the background does not have a very well defined frequency scaling. It appears thus, in the observations  at various frequencies, as an excess emission correlated between channels. The fluctuations of this background are expected to be significant at high galactic latitudes (where not masked by much stronger emissions from our own galaxy), and essentially at high frequencies (in the highest frequency channels of the Planck HFI.

\subsection{Point sources}

The ``point sources" component comprises all emissions from astrophysical objects such as radio galaxies, infrared galaxies, quasars, which are not resolved by the instruments used in CMB observations. For such sources, the issues are both their detection and the estimation of parameters describing them (flux at various frequencies, location, polarisation...), and specific methods are devised for this purpose. For diffuse component separation, they constitute a source of trouble. Usually, pixels contaminated by significant point source emission are blanked for diffuse component separation.


\section{Reduction of foreground contamination} \label{sec:cleaning}

The simplest way of avoiding foreground contamination consists in using prior information on emissions to reduce their impact on the data: by adequate selection of the region of observation, by masking some directions in the sky, by choosing the frequency bands of the instrument, or, finally, by subtracting an estimate of the contamination. All of these methods have been used widely in the context of CMB experiments.

\subsection{Selection of the region of observation}

Perhaps the most obvious solution to avoid contamination by foregrounds is to design the observations in such a way that the contamination is minimal. This sensible strategy has been adopted by ground--based and balloon--borne experiments observing only a part of the sky. In this case, CMB observations are made away from the galactic plane, in regions where foreground contamination from the galactic emissions is known to be small. The actual choice of regions of observation may be based on existing observations of dust and synchrotron emission at higher and lower frequencies, picking those regions where the emission of these foregrounds is known to be the lowest.

The drawback of this strategy is that the observations do not permit to estimate very well the level of contamination, nor the properties of the foregrounds.

\subsection{Masking}

For all-sky experiments, a strategy for keeping the contamination of CMB observations by foregrounds consists in masking regions suspected to comprise significant foreground emissions, and deriving CMB properties (in particular the CMB power spectrum) in the ``clean" region. The drawback of this strategy is that sky maps are incomplete.

Typically, for CMB observations, pixels contaminated by strong point sources (radio and infrared galaxies) are blanked, as well as a region containing the galactic plane. Such masks have been used in the analysis of WMAP data.

\subsection{Selection of the frequency bands of the instrument}

Of course, the selection of the frequency of observation to minimize the overall foreground contamination is a sensible option. For this reason, many CMB experiments aim at observing the sky around 70--100 GHz. Ground-based observations, however, need to take into account the additionnal foreground of atmospheric emission, which leaves as best windows frequency bands around 30 GHz, 90 GHz, 150 GHz, and 240 GHz.

\begin{figure} [thb]
\begin{center} 
\includegraphics[width=0.9\textwidth]{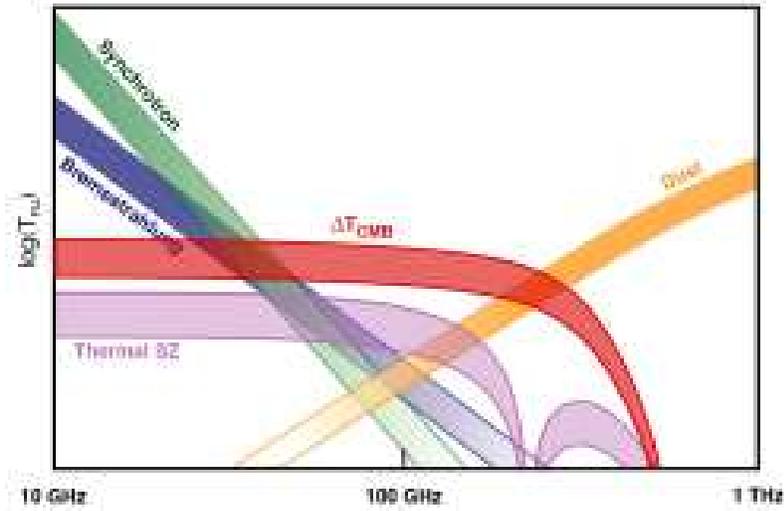} 
\caption{The frequency scaling of CMB and most relevant diffuse foregrounds, in Rayleigh-Jeans temperature, between 10 GHz and 1 THz. Depending on the relative amplitude of synchrotron, bremsstrahlung and dust emissions, the minimum of galactic foregrounds is somewhat below 100 GHz. 
Free-free emission decreases roughly as $\nu^{-2.1}$ and synchrotron as $\nu^{-3}$, while dust insreases as $\nu^2$.
The SZ effect is the major emission towards rich clusters, but is very localised. The thickness of the bands illustrates uncertainties as to the level of foregrounds, as well as uncertainties in the frequency scaling for synchrotron, free-free and dust emissions. Anomalous dust emission is not represented, due to our present lack of knowledge of the existence and nature of such a component.} 
\label{fig:foreground-freq-scaling}
\end{center} 
\end{figure}

Figure \ref{fig:foreground-freq-scaling} shows the expected typical frequency scalings for the major diffuse emission astrophysical components, including the CMB. For efficient component separation, CMB experiment, ideally, should comprise two or three channels around 70-100 GHz where CMB dominates, one channel around 217 GHz (the zero of the SZ effect), two channels at higher frequencies to monitor dust emission, and 3-4 channels at lower frequencies to monitor low frequency foregrounds.

Below 100 GHz, the present state of the art technology suggests the use of radiometers with high electron mobility transistor (HEMT) amplifiers, whereas above 100 GHz, low temperature bolometers provide a significantly better sensitivity that any other techniques. Typically, a single experiment uses one technology only. For Planck specifically, two different instruments have been designed to cover all the frequency range from 30 to 850 GHz.

\subsection{Foreground cleaning}

As a refinement to the above simple observational strategies, a first-order estimate of foreground contamination, based on observations made at low and high frequencies, can be subtracted from the observations. Depending on the accuracy of the model, the overall level of contamination can be reduced by a factor of a few at least, which permits to reduce the amount of cut sky.
This strategy, in particular, has been used by the WMAP team for the analysis of first year WMAP data \cite{2003ApJS..148...97B}.

Observations at low frequencies (10-40 GHz) can be used to map synchrotron emission and model its contribution in the 70-100 GHz range. Similar strategies can be used towards the high frequency side to model dust emission and subtract its contribution from CMB channels. For this purpose, models of emission as good as possible are needed, and the cleaning can be no better than the model used. There is always, therefore, a trade-off between a sophisticated model with simple correction methods (subtraction of an interpolation, simple decorrelation), and a simple model with sophisticated statistical treatments (multi-frequency filtering, independent component analysis). Which approach is best depends on a number of issues, and the answer is not completely clear yet.

\section{The linear model and system inversion} \label{sec:linear-system}

The most popular model of the observations for source separation in the context of CMB observations probably is the linear mixture. 

In this model, all components are assumed to have an emission which can be decomposed as the product of a spatial template independent of the frequency of observation, and of a spectral emission law which does not depend on the pixel. The total emission at frequency $\nu$, in pixel $p$, of a particular emission process $j$ is written as
$$
x_j(\nu,p) = a(\nu)s_j(p)
$$
or alternatively, in spherical harmonics space, 
$$
x_j(\nu,\ell m) = a(\nu)s_j(\ell m)
$$
Forgetting for the moment annoying details concerning the response of the instrument (beams, frequency bands, etc...) the observation with a detector is then:
$$
y_i(p) = \sum_j x_j(\nu_i,p) + n_i(p)
$$
where $n_i(p)$ is the contribution of noise for detector $i$. For a set of detectors, this can be recast in a matrix--vector form as
\begin{equation}
\vec{y}(p) = {\mA} \vec{s}(p) + \vec{n}(p)
\label{eq:linear-system}
\end{equation}
Here, $\vec{y}(p)$ represent the set of maps observed with all detectors detector, and $\vec{s}(p)$ are the unobserved components (one template map per astrophysical component). The mixing matrix $\mA$ which does not depend on the pixel for a simple linear mixture, has one column per astrophysical component, and one line per detector. 

If the observations are given in CMB temperature for all detectors, and if the detectors are properly calibrated, each element of the column of the mixing matrix corresponding to CMB is equal to 1.

The problem of component separation consists in inverting the linear system of equation
\ref{eq:linear-system}. Here we first concentrate on linear inversion, which consists in finding the ``best" possible matrix $\mW$ (such that $ \widehat {\vec{s}} = \mW {\vec{y}}$ is ``as good an estimator of $\vec{s}$ as possible").

\mypar{Covariances and multivariate power spectra\\}

In the following, a lot of use will be made of second order statistics of various sorts of data. In general, for a collection of maps ${\vec{x}}(p) = \{ x_i(p) \}$, the covariance will be noted as ${\mRx}(p,p')$, the elements of which are:
$$
R_{ij}(p,p')= \covar (x_i(p),x_j(p'))
$$
Alternatively, in harmonic space, we denote as ${\mRx}(\ell)$ the multivariate power spectrum of $\vec{x}$, i.e. the collection of matrices
$$
{\mRx}(\ell) = \langle \vec{x}(\ell,m) \vec{x}^\dagger (\ell,m) \rangle
$$ 
where the brackets $\langle . \rangle$
denote ensemble average, and the dagger $^\dagger$ denotes the transpose of the complex conjugate.
Such a power spectrum  is well defined only for stationary/isotropic random fields on the sphere 
for which $\langle \vec{x}(\ell,m) \vec{x}^\dagger (\ell,m) \rangle$ does not depend on $m$.

\subsection{Simple inversion}
If $\mA$ is square and non singular, in absence of any additional information, then the inversion is obtained by
\begin{equation}
\mW = \mA^{-1}
\end{equation}
and we have
\begin{equation}
\widehat {\vec{s}}  = \mA^{-1} {\vec{y}} =  {\vec{s}} + \mA^{-1} {\vec{n}}
\end{equation}
Note that because of the remaining noise term, this inversion is not always the best solution in terms of residual error, in particular in the poor signal to noise regimes. For instance, if we have two measurements of a mixture of CMB + thermal dust in a clean region of the sky (low foregrounds), one of which, at 150 GHz, is relatively clean, and the other, at 350 GHz, quite poor because of high level noise, then it may be better to use the 150 GHz as the CMB template (even with some dust contamination), rather than to invert the system, subtracting little dust and adding a large amount of noise.

In terms of residual foreground contamination however (if the criterion is to reject astrophysical signals, whatever the price to pay in terms of noise), the only solution here is matrix inversion. The solution is unbiased, but may be noisy.

Note that an ILC method would produce a different solution, possibly slightly biased (as discussed in \ref{sub:comments-ILC}), but possibly better in terms of signal to noise ratio of the end product.

This solution can be applied if the full matrix $\mA$ is known (not only the column of the component of interest, i.e. the CMB), without further prior knowledge of the data.

\subsection{Inversion of a redundant system using the pseudo inverse}
If there are more observations than components, but nothing is known about noise and signal levels, then the inversion is obtained by
\begin{equation}
\mW = \left [ \mA^\dagger \mA \right ] ^{-1} \mA^\dagger 
\end{equation}
and we have
\begin{equation}
\widehat {\vec{s}}  =\left [ {\mA}^\dagger {\mA} \right ] ^{-1} {\mA}^\dagger  {\vec{y}} =  {\vec{s}} + \left [ {\mA}^\dagger {\mA} \right ] ^{-1} \mA^\dagger {\vec{n}}
\end{equation}
Again, this estimator is unbiased, but may contain a large amount of noise and may not be optimal in terms of signal to noise ratio. All the comments made in the previous paragraph hold as well for this solution. 

Note that there is no noise-weighting here, so that one single very bad channel may contaminate significantly all the data after inversion. It is therefore not a very good idea to apply this estimator with no further thoughts.

Note that, again, this solution can be implemented without any further knowledge about signal and noise -- only the entries of the mixing matrix for all components are needed.

\subsection{A noise-weighted scheme: the Generalised Least-Square solution}
Let us now assume that we know something additional about the noise, namely, its second order statistics. These are described by noise correlation matrices in real space, or alternatively by noise power spectra in Fourier (for small maps) or in harmonic (for all-sky maps) space.

We denote as $\mRn$ the noise correlation matrix and assume, for the time being, that the noise for each detector $i$ is a realization of a random gaussian field, the generalised (or global) least square (GLS) solution of the system of equation \ref{eq:linear-system} is:
\begin{equation}
{\mW} = \left [ {\mA}^\dagger {\mRn}^{-1} {\mA} \right ] ^{-1} \mA^\dagger {\mRn}^{-1} 
\label{eq:GLS-matrix}
\end{equation}
and we have
\begin{equation}
\widehat {\vec{s}}  =\left [ {\mA}^\dagger {\mRn}^{-1} {\mA} \right ] ^{-1} {\mA}^\dagger  {\mRn}^{-1} {\vec{y}} =  {\vec{s}} + \left [ {\mA}^\dagger {\mRn}^{-1} {\mA} \right ] ^{-1} {\mA}^\dagger {\mRn}^{-1} {\vec{n}}
\end{equation}
Again, the solution is unbiased. Altough there remains a noise contribution, this is the solution yielding the minimum variance error map for a deterministic signal (in contrast with the Wiener solution below, which optimises the variance of the error when the signal is stochastic, i.e. assumed to be a random field). It is also the best linear solution in the limit of large signal to noise ratio. 

This solution is also theoretically better than the ILC when the model holds, but the price to pay is the need for more prior knowledge about the data (knowledge of the mixing matrix and of noise covariance matrices or power spectra). If that knowledge is unsufficient, one has to design methods to get it from the data itself. Such ``model learning" methods will be discussed in section \ref{sec:blind}.

\subsection{The Wiener solution}
The Wiener filter \cite{1949WienerFilter} has originally been designed to filter time series in order to suppress noise, but has been extended to a large variety of applications since then. Wiener's solution requires additional information regarding the spectral content of the original signal and the noise. Wiener filters are characterized by the following:
\begin{itemize}
\item Both the noise and the signal are considered as stochastic processes with known spectral statistics (or correlation properties) -- contrarily to the GLS method which considers the noise only to be stochastic, the signal being deterministic,
\item The optimization criterion is the minimum least square error,
\item The solution is linear.
\end{itemize}
In signal processing, a data stream $y(t) = s(t)+n(t)$ assumed to be a noisy measurement of a signal $s$ can be filtered for denoising as follows: in Fourier space, each mode $y(f)$ of the data stream is weighted by a coefficient 
$$
W(f) = \frac{S(f)}{S(f)+N(f)}
$$
where $S(f) = \langle |s(f)|^2 \rangle$ and $N(f) = \langle |n(f)|^2 \rangle$ are ensemble averages of the square moduli of the Fourier coefficients of the stochastic processes $s$ and $n$.

In the limit of very small noise level $N(f) \ll S(f)$, the Wiener filter value is $W(f)=1$, and the filter does not change the data. In the limit of very poor signal to noise  $S(f) \ll N(f)$, the filter suppresses the data completely, because that mode adds noise to the total data stream, and nothing else.

It can be shown straightforwardly that the Wiener filter minimizes the variance of the error of the signal
estimator $\widehat s(f) = W(f) y(f)$ (so that $\langle \int_f | \widehat s(f) - s(f)|^2 {\rm{d}}f \rangle$ is minimal).

The Wiener solution can be adapted for solving our component separation problem, provided the mixing matrix $\mA$ and the second order statistics of the components and of the noise are known \cite{1996MNRAS.281.1297T,1999NewA....4..443B} as:

\begin{equation}
{\mW}^{(1)} = \left [ {\mA}^\dagger {\mRn}^{-1} {\mA} + {\mRs}^{-1}  \right] ^{-1} {\mA}^\dagger {\mRn}^{-1} 
\label{eq:wiener-matrix}
\end{equation}
where ${\mRs}$ is the correlation matrix of the sources (or power spectra of the sources, in the Fourier or harmonic space), and $\mRn$ the correlation matrix of the noise. The superscript $(1)$ is used to distinguish two forms of the Wiener filter (the second is given later in this section). 

An interesting aspect of the Wiener filter is that:
\begin{eqnarray}
\widehat {\vec{s}}  & = & \left [ {\mA}^\dagger {\mRn}^{-1} {\mA} + {\mRs}^{-1} \right ] ^{-1} {\mA}^\dagger  {\mRn}^{-1} {\vec{y}} \nonumber \\
& =  & \left [ {\mA}^\dagger {\mRn}^{-1} {\mA} + {\mRs}^{-1} \right  ]^{-1} {\mA}^\dagger  {\mRn}^{-1}  {\mA} {\vec{s}} \nonumber \\
& & + 
\left [ {\mA}^\dagger {\mRn}^{-1} {\mA} + {\mRs}^{-1} \right ] ^{-1} {\mA}^\dagger {\mRn}^{-1} {\vec{n}}
\end{eqnarray}
The matrix in front of $\vec{s}$ is not the identity, and thus the Wiener filter does not give an unbiased estimate of the signals of interest. Diagonal terms can be different from unity. In addition, non-diagonal terms may be non-zero, which means that the Wiener filter allows some residual foregrounds to be present in the final CMB map -- the objective being to minimise the variance of the residuals, irrespective of whether these residuals originate from instrumental noise or from astrophysical foregrounds.

As noted in \cite{1996MNRAS.281.1297T}, the Wiener solution can be ``debiased" by multiplying the Wiener matrix by a diagonal matrix removing the impact of the filtering. The authors argue that for the CMB this debiasing is desirable for subsequent power spectrum estimation on the reconstructed CMB map. Each mode of a given component is divided by the diagonal element of the Wiener matrix for that component and that mode. This, however, destroys the minimal variance property of the Wiener solution, and can increase the noise very considerably. There is an incompatibility between the objective ofobtaining a minimum variance map, and the objective of obtaining an unbiased map which can be used directly to measure the power spectrum of the CMB. There is no unique method for both. 

Before moving on, it is interesting to check that the matrix form of the Wiener filter given here reduces to the usual form when there is one signal only and when the matrix $\mA$ reduces to a scalar equal to unity. In that case, the Wiener matrix $\mW$ of equation \ref{eq:wiener-matrix} reduces to
$$
W(f) = [1/S(f) + 1/N(f)]^{-1} / N(f) = S(f)/[N(f)+S(f)]
$$
where $S$ and $N$ are the signal and noise power spectra, and we recover the classical Wiener formula.

\mypar{Two forms of the Wiener Filter\\}

In the literature, another form can be found for the Wiener filter matrix:
\begin{equation}
{\mW}^{(2)} = {\mRs} {\mA}^\dagger \left [ {\mRn} + {\mA} {\mRs} {\mA}^\dagger \right] ^{-1}
\label{eq:wiener-matrix2}
\end{equation}
It can be shown straightforwardly that if the matrices 
$${\mM}_1 = \left [ {\mA}^\dagger {\mRn}^{-1} {\mA} + {\mRs}^{-1}  \right]$$ 
and 
$${\mM}_2 = \left [ {\mRn} + {\mA} {\mRs} {\mA}^\dagger \right] $$ 
are regular, then the forms of equation \ref{eq:wiener-matrix} and \ref{eq:wiener-matrix2} are equivalent (simply multiply both forms by ${\mM}_1$ on the left and ${\mM}_2$ on the right, and expand).

It may seem that the form of equation \ref{eq:wiener-matrix2} is more convenient, as it requires only one matrix inversion instead of three. Each form, however, presents specific advantages or drawbacks, which appear clearly in the high signal to noise ratio (SNR) limit, and if power spectra of all signals are not known.

\mypar{The high SNR limit\\}
The two above forms of the Wiener filter are not equivalent in the high SNR limit. In this regime, equation 
\ref{eq:wiener-matrix} yields in the limit
$${\mW}_{\rm{limit}}^{(1)}  = \left [ {\mA}^\dagger {\mRn}^{-1} {\mA} \right] ^{-1} {\mA}^\dagger {\mRn}^{-1}$$
which is the
GLS solution of equation \ref{eq:GLS-matrix}, and depends only on the noise covariance matrix, whereas equation \ref{eq:wiener-matrix2} tends to 
$${\mW}_{\rm{limit}}^{(2)} = {\mRs} {\mA}^\dagger \left [{\mA} {\mRs} {\mA}^\dagger \right] ^{-1}$$ 
which depends only on the covariance of the signal. Therefore, some care should be taken when applying the Wiener filter in the high SNR ratio regime, when numerical roundup errors may cause problems.  

Note that if $\left [ {\mA}^\dagger{\mA} \right]$ is regular, then 
\begin{eqnarray}
{\mW}_{\rm{limit}}^{(2)} & = & {\mRs} {\mA}^\dagger \left [{\mA} {\mRs} {\mA}^\dagger \right] ^{-1} \nonumber \\
& = & \left [ {\mA}^\dagger{\mA} \right]^{-1} \left [ {\mA}^\dagger{\mA} \right] {\mRs} {\mA}^\dagger \left [{\mA} {\mRs} {\mA}^\dagger \right] ^{-1} \\
& = & \left [ {\mA}^\dagger{\mA} \right]^{-1} {\mA}^\dagger
\end{eqnarray}
and the limit is simply the pseudo inverse of matrix $\mA$, without any noise weighting. Of course, when there is no noise at all, ${\mW}^{(1)} $ can not be implemented at all, and the Wiener solution is pointless anyway.

\mypar{What if some covariances are not known?\\}
It is interesting to note that even if the covariance matrix (or equivalently multivariate power spectrum) $\mRs$ of all sources is not known, it is still possible to implement an approximate Wiener solution if the maps of observations are large enough to allow a good estimate of the covariance matrix of the observations.

If ${\vec{y}} = {\mA}{\vec{s}} + {\vec{n}}$ and if the noise and the components are independent, the covariance $\mRy$ of the observations is of the form:
$$
{\mRy} = {\mRn} + {\mA} {\mRs} {\mA}^\dagger
$$
Therefore, form 2 of the Wiener filter can be recast as:
\begin{equation}
{\mW}^{(2)} = {\mRs} {\mA}^\dagger \left [ {\mRy} \right] ^{-1}
\end{equation}
If all components are decorrelated, the matrix $\mRs$ is diagonal. For the implementation of a Wiener solution for one single component (e.g. CMB), only the diagonal element corresponding to the CMB (i.e. the power spectrum $C_\ell$ of the CMB) is needed, in addition to the multivariate power spectrum of the 
observations $\mRy$. The latter can be estimated directly using the observations.

\subsection{Comment on the various linear inversion solutions}

The above four linear solutions to the inversion of the linear system of equation \ref{eq:linear-system} have been presented by order of increasing generality, increasing complexity, and increasing necessary prior knowledge. The various solutions are summarised in table \ref{tab:linear-solutions}. Three comments are necessary. 

Firstly, we note that the Wiener solutions require the prior knowledge of the covariance matrices (or equivalently power spectra) of both the noise and the signal. For CMB studies, however, the measurement of the power spectrum of the CMB field is precisely the objective of the observations. Then, the question of whether the choice of the prior on the CMB power spectrum biases the final result or not is certainly of much relevance. For instance, the prior assumption that the power spectrum of the CMB is small in some $\ell$ range will result in filtering the corresponding modes, and the visual impression of the recovered CMB will be that indeed there is little power at the corresponding scales. For power spectrum estimation on the maps, however, this effect can be (an should be) corrected, which is always possible as the effective filter induced by the Wiener solution is known (for an implementation in harmonic space, it is equal for each mode $\ell m$, for each component, to the corresponding term of the diagonal of $\mW_{\ell m} \mA$). In section \ref{sec:smica}, a solution will be proposed for estimating first on the data themselves all the relevant statistical information (covariance matrices and frequency scalings), and then using this information for recovering maps of the different components.

Secondly, we should emphasise that the choice of a linear solution should be made with a particular objective in mind. If the objective is to get the best possible map in terms of least square error, then the Wiener solution is the best solution if the components are Gaussian. The debiased Wiener is not really adapted to any objective in particular. The GLS solution is the best solution if the objective is an unbiased reconstruction with no filtering and no contamination. In practice, it should be noted that small uncertainties on $\mA$ result in errors (biases and contamination) even for the GLS solution.

As a third point, we note that it can be shown straightforwardly that for Gaussian sources and noise, the Wiener solution maximises the posterior probability $P(\vec{s}|\vec{y})$ of the recovered sources given the data. From Bayes theorem, the posterior probability is the product of the likelihood $P(\vec{y}|\vec{s})$ and the prior $P(\vec{s})$, normalised by the evidence $P(\vec{y})$. The normalising factor does not depend on $s$. We can write:
\begin{eqnarray}
P(\vec{s}|\vec{y}) & \propto & {\rm exp} \left [ - {(\vec{y}-\mA \vec{s})^\dagger \mRn ^{-1}(\vec{y}-\mA \vec{s})} \right ]
{\rm exp} \left [-  \vec{s}^\dagger \mRs ^{-1} \vec{s} \right ] \nonumber \\
{\rm log}(P(\vec{s}|\vec{y}) ) & = & - \left [ {(\vec{y}-\mA \vec{s})^\dagger \mRn ^{-1}(\vec{y}-\mA \vec{s})} \right ] - 
\left [ \vec{s}^\dagger \mRs ^{-1} \vec{s} \right ] + {\rm const.} 
\end{eqnarray}
where ${\rm exp} \left [ - \vec{s}^\dagger \mRs ^{-1} \vec{s} \right ] $ is the Gaussian prior for $s$.
The requirement that 
$$
\frac{\partial}{\partial \vec{s}} {\rm log}(P(\vec{s}|\vec{y}) ) =  0
$$
implies
\begin{eqnarray}
- \mA^\dagger \mRn ^{-1}(\vec{y}-\mA \vec{s}) + \mRs ^{-1} \vec{s} & = & 0 \nonumber \\
\mA^\dagger \mRn ^{-1}\mA \vec{s} + \mRs ^{-1} \vec{s} & = & \mA^\dagger \mRn ^{-1}\vec{y} \nonumber \\
\left [ \mA^\dagger \mRn ^{-1}\mA + \mRs ^{-1} \right ]  \vec{s} & = & \mA^\dagger \mRn ^{-1}\vec{y}
\end{eqnarray}
and thus we get the solution $\mW^{(1)}$. In section \ref{sec:maxent}, we will dicuss the case where the Gaussian prior is replaced by an entropic prior, yielding yet another solution for $s$.

\begin{table}[htpb]
\begin{center}
\begin{tabular}{|c|c|c|c|}
\hline
Solution & ${\mW} =$ & \begin{minipage}{0.12\textwidth} \begin{center} \vspace{.3cm}Required prior knowledge \vspace{.3cm} \end{center} \end{minipage} & Comments \\
\hline
Inverse & $\mA^{-1}$ & $\mA$ & \begin{minipage}{0.25\textwidth} \vspace{.3cm} When there are as many channels of observation as components. Unbiased, contamination free. \vspace{.3cm} \end{minipage}\\
\hline
Pseudo--inverse & $\left [ \mA^\dagger \mA \right ] ^{-1} \mA^\dagger $ & $\mA$ & \begin{minipage}{0.25\textwidth} \vspace{.3cm} When there are more channels of observation than components. Unbiased, contamination free. \vspace{.3cm} \end{minipage}\\
\hline
GLS & $\left [ {\mA}^\dagger {\mRn}^{-1} {\mA} \right ] ^{-1} \mA^\dagger {\mRn}^{-1} $ & $\mA$  and ${\mRn}$ & \begin{minipage}{0.25\textwidth} \vspace{.3cm} Minimises the variance of the error for deterministic signals. Unbiased, contamination free.\vspace{.3cm} \end{minipage}\\
\hline
Wiener 1 & $\left [ {\mA}^\dagger {\mRn}^{-1} {\mA}  + {\mRs}^{-1} \right ] ^{-1} \mA^\dagger {\mRn}^{-1} $ & $\mA$, ${\mRn}$  and ${\mRs}$ & \begin{minipage}{0.25\textwidth} \vspace{.3cm} Minimises the variance of the error for stochastic signals. Biased, not free of contamination. Tends to the GLS solution in the limit of high SNR. \vspace{.3cm} \end{minipage}\\
\hline
Wiener 2 & ${\mRs} {\mA}^\dagger \left [ {\mRn} + {\mA} {\mRs} {\mA}^\dagger \right] ^{-1}$ & $\mA$, ${\mRn}$  and ${\mRs}$ & \begin{minipage}{0.25\textwidth} \vspace{.3cm} Equivalent to Wiener 1. Tends to the pseudo inverse in the limit of high SNR. \vspace{.3cm} \end{minipage}\\
\hline
Debiased Wiener& $\Lambda {\mRs} {\mA}^\dagger \left [ {\mRn} + {\mA} {\mRs} {\mA}^\dagger \right] ^{-1}$ & $\mA$, ${\mRn}$  and ${\mRs}$ & \begin{minipage}{0.25\textwidth} \vspace{.3cm} The diagonal matrix $\Lambda$, inverse of the diagonal of  $\mW \mA$ where $\mW$ is the Wiener solution, removes for each mode the filtering effect of the Wiener filter. Unbiased, but not contamination free. \vspace{.3cm} \end{minipage}\\
\hline
\end{tabular}
\caption{Summary of linear solutions to component separation when the mixing matrix ${\mA}$ is known.}
\label{tab:linear-solutions}
\end{center}
\end{table}

\subsection{Pixels, harmonic space, or wavelets?}

The simple inversion of $\mA$ using the inverse or pseudo-inverse can be implemented equivalently with any representations of the maps, in pixel domain, harmonic space, or on any decomposition of the observations on a set of functions as, e.g., wavelet decompositions \cite{2005moudden}. The result in terms of separation is independent of this choice, as far as the representation arises from a linear transformation.

If all sources and signals are Gaussian random fields, the same is true for GLS or Wiener inversions, provided all the second order statistics are properly described by the covariance matrices $\mRn$ and $\mRs$.

These covariance matrices, in pixel space, take the form of the set of covariances:
$$ 
\mRn = \{ R_{n_i n_j}(p_i,p_j)\}
$$
where 
$$R_{n_i n_j}(p_i,p_j) = \langle  n_i(p_i) n_j(p_j) \rangle$$
Similarly, in harmonic space, we have:
$$
\mRn = \{ R_{n_i n_j}(\ell_i,m_i,\ell_j,m_j)\}
$$
where 
$$
R_{n_i n_j}(\ell_i,m_i,\ell_j,m_j) = \langle  n_i(\ell_i,m_i) n_j(\ell_j,m_j) \rangle
$$

If the number of pixels is large, if we deal with several sources and many channels at the same time (tens today, thousands in a few years), the implementation of the GLS or Wiener solution may be quite demanding in terms of computing. For this reason, it is desirable to implement the solution in the space where matrices are the most easy to invert.

For stationnary Gaussian random fields, harmonic space implementations are much easier then direct space implementations, because the covariance between distinct modes vanish, so that
$$
R_{n_i n_j}(\ell_i,m_i,\ell_j,m_j) = \langle  n_i(\ell_i,m_i) n_j(\ell_i,m_i) \rangle \delta_{\ell_i \ell_j} \delta_{m_i m_j}
$$
The full covariance matrix consists in a set of independent covariance matrices (one for each mode), each of which is a small matrix of size $N_{\rm channels} \times N_{\rm channels}$ for $\mRn$, and of size $N_{\rm sources} \times N_{\rm sources}$ for $\mRs$.

\subsection{Annoying details}

Under the assumption that the response of each detector $i$ in the instrument can itself be decomposed in the product of a spectral response $h_i(\nu)$ and a frequency independent symmetrical beam $B_i$, the contribution of component $j$ to the observation obtained with detector $i$ is:
$$
y_{ij}(\ell m) = \left [ \int_\nu h_i(\nu)a(\nu) {\rm d} \nu \right ] B_{i,\ell} \, s_j(\ell m)
$$
where $B_{i,\ell}$ are the coefficients of the expansion of the symmetric beam of detector $i$ on Legendre polynomials.

The mixing matrix of this new linear model is seen to include a band integration, assumed to first order to be independent of $\ell$, a the effect of a beam, which depends on $\ell$. Both can be taken into account in a linear inversion, if known a priori.

\section{The Maximum Entropy Method} \label{sec:maxent}

The Wiener filter provides the best (in terms of minimum-variance, or maximum likelihood) estimate of the component maps if two main asumptions hold. Firstly, the observations should be a linear mixture of distinct emissions. Secondly, the components and the noise should be (possibly correlated) Gaussian stationnary random processes.

Unfortunately, the sky is known to be neither Gaussian, nor stationnary, with the possible exception of the CMB itself. Is this critical?

The Maximum Entropy Method (MEM) of component separation is a method which inverts the same linear system of component mixtures, but assumes non-gaussian probability distributions  \cite{1998MNRAS.300....1H}.

\subsection{Maximum Entropy}

The concept of entropy in information theory has been introduced by Shannon in 1948 \cite{shannon-entropy}. The entropy of  a discrete random variable $X$ on a finite set of possible values $\{ x_i \}$ with probability distribution function $p(x_i) = p(X=x_i)$, is defined as:
\begin{equation}
H(X) = - \sum_{i=1}^{N} p(x_i) \log p(x_i)
\label{eq:entropy-discrete}
\end{equation}

The principle of maximum entropy is based of the idea that whenever there is some choice to be made about the distribution function of the random variable $X$, one should choose the least informative option possible. Entropy measures the amount of information in a probability distribution, and entropy maximisation is a way of achieving this.

For instance, in the absence of any prior information, the probability distribution which maximises the entropy of equation \ref{eq:entropy-discrete} is a distribution with uniform probability, $p(x_i)=1/N$, i.e.  the least informative choice of a probability distribution on the finite set $\{ x_i \}$, where all outcomes are equally likely. This is the most natural choice if nothing more is said about the probability distribution.

In the opposite, a most informative choice would be a probability which gives a certain result, (for instance always $X=x_1$). This is a probability distribution which minimizes entropy.

In the continuous case where $X$ can achieve any real value $x$ with probability density $p(x)$, entropy can be defined as:
\begin{equation}
H(X) = - \int_{-\infty}^{\infty} p(x) \log p(x) \, dx
\label{eq:entropy-continuous}
\end{equation}
Of course, maximum entropy becomes really useful when there is also additional information available. In this case, the entropy must be maximized within the constraints given by additional information. 

For instance, the maximum entropy distribution of a real random variable of mean $\mu$ and variance $\sigma^2$ is the normal (Gaussian) distribution :
$$
p(x) = \frac{1}{2 \pi \sigma} \exp \left [ - \frac{(x-\mu)^2}{2 \sigma ^2}  \right ]
$$
For this reason, in absence of additional information about the probability distribution of a random variable of known mean and variance, it is quite natural, according to the Maximum Entropy principle, to assume a Gaussian distribution -- which maximises the entropy, and hence corresponds to the least informative choice possible.

An other useful example is the maximum entropy distribution of a real positive random variable of mean $\mu$, which is the exponential distribution :
$$
p(x) = \frac{1}{\mu} \exp (-x / \mu)
$$

\subsection{Relative entropy}

In fact, the differential entropy of equation \ref{eq:entropy-continuous} has an unpleasant property. It is not invariant under coordinate transformations (on spaces with more than one dimension).

The definition of \emph{relative entropy} (or Kullback-Leibler divergence) between two distributions solves the issue. It can be interpreted as a measure of the amount of additional information one gets from knowing the actual (true) probability distribution $p(x)$, instead of an imperfect model $m(x)$, and is given by:
\begin{equation}
D_{\rm KL}(p||m) = \int_{-\infty}^{\infty} p(x) \log \frac{p(x)}{m(x)} \, dx
\end{equation}

Later in this paper (in section \ref{sec:smica}), we will make use of the Kullback-Leibler divergence for measuring the ``mismatch" between two positive matrices $\mRone$ and $\mRtwo$. It will actually correspond to the KL divergence between two Gaussian distributions with covariance matrices  $\mRone$ and $\mRtwo$. 


The relative entropy is invariant under coordinate transformations (because both the ratio $p(x)/m(x)$ and $p(x)dx$ are invariant under coordinate transformations).

\subsection{Component separation with the MEM}

In principle, replacing the Gaussian prior by some other prior is perfectly legitimate. In practice, the choice of such a prior is not obvious, as the full statistical description of a complex astrophysical component is difficult to apprehend..

Following the maximum entropy principle, one may decide to use as a prior the distribution which maximises the entropy given a set of constraints. If the constraints are the value of the mean, and the variance, then the maximum entropy prior is the Gaussian prior.

Hobson and collaborators, in their MEM paper \cite{1998MNRAS.300....1H}, argue that based on the maximum entropy principle, an appropriate prior for astrophysical components $s$ is

\begin{equation}
p(s) = \exp \left [ - \alpha S_c(\vec{s},\vec{m}_u,\vec{m}_v)   \right ]
\label{eq:entropic-prior}
\end{equation}
with 
$$
S_c(\vec{s},\vec{m}_u,\vec{m}_v) = \sum_{j=1}^L \left \{ \psi_j - m_{uj} - m_{vj} - 
s_j \ln \left [  \frac{\psi_j + s_j }{2m_{uj}}  \right ]    \right \}
$$
where $\psi_j = [{s_j}^2 + 4m_{uj}m_{vj}]^{1/2}$, and where $\vec{m}_u$ and $\vec{m}_v$ are models of two positive additive distributions (which are not clearly specified) used to represent the astrophysical components.

A derivation for this is given in \cite{1998MNRAS.298..905H}, but the connection to entropy is not direct. In particular, the definition of entropy does not require the \emph{values} of the random variables to be positive, but their \emph{probability densities}, which makes the discussion unconvincing. 

Pragmatically, the choice for the prior of equation \ref{eq:entropic-prior} seems to be validated a posteriori by the performance of the separation, which is not worse (and actually better for some of the components) than that obtained with the Wiener filter. It is not likely to be optimal, however, because the non-stationnarity of components implies correlations in the harmonic domain, which are not fully taken into account in the MEM implementation.

The maximisation of the posterior probability (and hence of the product of the likelihood and the prior), is done with a dedicated fast maximisation algorithm. We refer the reader to the relevant papers for additional details \cite{1998MNRAS.298..905H,2002MNRAS.336...97S}. 

This method has been applied to the separation of components in the COBE data \cite{2004MNRAS.351..515B}.

\subsection{Comments about the MEM}

Although entropy has a clear meaning in terms of information content in the discrete case (e.g. it defines the minimum number of bits necessary to represent a sequence), there is no such interpretation in the continuous case. Entropy maximisation, understood as minimising the amount of arbitrary information
in the assumed distribution, hence, is not very clearly founded for continuous images.

The ``principle" of maximum entropy, as the name indicates, is not a theorem, but a reasonable recipe which seems to work in practice. In the context of the CMB, there is no guarantee that it is optimal, among all non-linear solutions of the mixing system.
MEM outperforms the Wiener filter solution for some components in particular because the entropic prior of Hobson and Lasenby allows heavier tails than the Gaussian prior. Other priors however, based on a physical model of the emissions, might well perform even better in some cases. This question remains as an open problem in the field.

\section{ICA and Blind source separation} \label{sec:blind}

\subsection{About blind separation}

The term ``blind separation'' refers to a fascinating possibility: if
the components of a linear mixture are \emph{statistically
  independent}, they can be recovered even if the mixing matrix
$\mathbf{A}$ is unknown \textit{a priori}.  In essence, this is
possible because statistical independence is, at the same time, a
strong mathematical property and, quite often, a physically plausible
one.

There is an obvious and strong motivation for attempting blind
component separation: allowing underlying components to be recovered
blindly makes it possible to analyze multi-detector data with limited,
imperfect, or even outright missing knowledge about the emission laws
of the components.  Even better, one can process data without knowing
in advance which components might be ``out there''.  Hence, the blind
approach is particularly well suited for exploratory data analysis.

In the last fifteen years, blind component separation has been a very
active area of research in the signal processing community where it
goes by the names of ``blind source separation'' (BSS) and
``independent component analysis'' (ICA).
This section outlines the principles underlying some of the best known
approaches to blind source separation.  There is not a single best
approach because there is not a unique way in which to express
statistical independence on the basis of a finite number of samples.

\subsection{Statistical independence}

This section explains why blind component separation is possible in
the first place.  For the sake of exposition, the main ideas are
discussed in the simplest case: there is no observation noise and
there are as many ``channels'' as underlying components.  Thus the
model reduces to
\begin{displaymath}
  \by(p) = \A \bs(p)
\end{displaymath}
where $\A$ is an $\nbs\times\nbs$ matrix and we are looking for an
$\nbs\times\nbs$ matrix ``separating matrix'' $\W$.  Of course, if the
mixing matrix $\A$ is known, there is little mystery about separation:
one should take $\W=\A\inv$ and be done with it.

If nothing is known about $\A$ but the components are known (or
assumed) to be statistically independent, the idea is to determine
$\W$ in such a way that the entries of vector $\W\by$ are independent
(or as independent as possible).  In other words, the hope is that by
restoring independence, one would restore the components themselves.
Amazingly enough, this line of attack works.  Even better, under
various circumstances, it can be shown to correspond to maximum
likelihood estimation and there is therefore some statistical
optimality to it\ldots provided the hypothesis of statistical
independence is expressed vehemently enough.  

Note however that no matter the amount of statistical ingenuity thrown
at blind component separation, there is no hope to recover completely
the mixing matrix (or equivalently: the components).  This is because
a scalar factor can always be exchanged between each entry of $\bs$
and the corresponding column of $\A$ without changing what the model
predicts (\textit{i.e.} the value of the product $\A\bs$) and without
destroying the (hypothetical) independence between the entries of
$\bs$.  The same is true of a renumbering of the columns of $\A$ and
of the entries of $\bs$.  In other words, blind recovery is possible
only up to rescaling and permutation of the components.  In many
applications, this will be ``good enough''.  If these indeterminacies
have to be fixed, it can be done only by imposing additional
constraints or resorting to side information.

For any possible choice $\W$ of a candidate separating matrix, denote
\begin{displaymath}
  \bx(p)=\W\by(p)
\end{displaymath}
the corresponding vector of candidate components.  If $\W=\A\inv$ then
the entries of $\bx$ are independent (since, in this case
$\bx(p)=\bs(p)$).  Under which circumstances would the \emph{converse}
true?  Whenever the converse is true, it will be possible to recover
the sources by looking for the linear transform $\W$ which makes them
independent.  Hence, we have a blind separation principle: to separate
components, make them independent.

\subsection{Correlations}

The main difficulty in blind source separation is to define a measure
of independence.  The problem is that the simple decorrelation
condition\footnote{Here, as in the rest of this section, all signals
are assumed to have zero mean.}  between any two candidate components:
\begin{equation}\label{eq:simplecorr}
  \frac1\nbe \sum_{p=1}^\nbe x_i(p)x_j(p) =0
  \quad\mbox{for}\quad 1\leq i\neq j\leq\nbs  .
\end{equation}
\emph{does not cut it}. 
This is in fact obvious from the fact that this decorrelation
condition between $x_i$ and $x_j$ is \emph{symmetric}.  Hence
decorrelation provides only $\nbs(\nbs-1)/2$ constraints while
$\nbs^2$ constraints are needed to determine $\W$ uniquely.
Therefore, more expressive forms of independence must be used.  Two
main avenues are possible: non-linear correlations and localized
correlations, as described next.

\paragraph{Non-linear correlations}
The ``historical approach'' to blind separation has been to determine
a separating matrix $\W$ in order to obtain ``non-linear
decorrelations'' \emph{i.e.}
\begin{equation}\label{eq:esteqng}
  \frac1\nbe \sum_{p=1}^\nbe \psi_i(x_i(p)) \; x_j(p) = 0
  \quad\mbox{for}\quad 1\leq i\neq j\leq\nbs
\end{equation}
where functions $\psi_1, \ldots,\psi_\nbs: R\mapsto R$ are
\emph{non-linear} functions (more about choosing them below).  By
using non-linear functions, symmetry is broken and the required number
of constraints is obtained, namely $\nbs(\nbs-1)$ (with $\nbs$
additional arbitrary constraints, needed for fixing the scale of each
component.

\paragraph{Localized correlations}
Another approach is to look for ``localized decorrelation'' in the
sense of solving
\begin{equation}\label{eq:esteqloc}
  \frac1\nbe \sum_{p=1}^\nbe \frac {x_i(p)}{\sigma_{ip}^2} x_j(p) = 0
  \quad\mbox{for}\quad 1\leq i\neq j\leq\nbs
\end{equation}
where for each component $i$, a sequence $\{ \sigma_{ip}\}_{p=1}^\nbe$
of positive number must be defined (more about this soon).  Again,
blind identification is possible because symmetry is broken, provided
no two sequences of $\sigma$'s are proportional.  

\paragraph{Maximum likelihood}

Why using the particular proposals~(\ref{eq:esteqng})
or~(\ref{eq:esteqloc}) as extended decorrelation conditions rather
than any other form, possibly more complicated?  
One reason is that reasonable algorithms exist for computing the $\W$
such that $\bx=\W\by$ is a solution of (\ref{eq:esteqng}) or
(\ref{eq:esteqloc}).  
Another, more important reason is that these two conditions actually
characterize the maximum likelihood estimate of $\W$ in simple and
well understood models.  Because of this, we can understand what the
algorithm does and we have guidance for choosing the non-linear
functions $\psi_i$ in condition~(\ref{eq:esteqng}) or the varying
variance profiles $\sigma_{iq}^2$ in condition~(\ref{eq:esteqloc}) as
stated next.

\textbf{Non linear correlations.}  
Assume that each component $\{s_i(p)\}$ is modeled as having all
pixels independently and identically distributed according to some
probability density $p_i$.  In this model, the most likely value of
$\A$ given the observations has for inverse a matrix $\W$ such that
condition~(\ref{eq:esteqng}) holds with $\psi_i=-p_i'/p_i$.  Hence, if
the model is true (or approximately true), the non linear-function
appearing in condition~(\ref{eq:esteqng}) should be taken as minus the
derivative of the log-density of $s_i(p)$.  For a Gaussian
distribution $p_i$, the corresponding function $\psi_i$ is linear:
here, the necessary non-linearity of $\psi_i$ corresponds to the non
Gaussianity of the corresponding component.

\textbf{Localized correlations.} 
Alternatively, one may model each component $\{s_i(p)\}$ as having all
pixels independently and \emph{normally} distributed with zero-mean
and ``local'' variance $\sigma_{ip}^2$.  Then, in this model, the
likeliest value of $A$ given the observations has for inverse a matrix
$\W$ such that $\bx=\W\by$ satisfies condition~(\ref{eq:esteqloc}).

\subsection{ICA in practice}

For the simple noise-free setting under consideration (the noisy case
is addressed in next section), the algorithmic solutions depend on the
type of decorrelation one decides to use.

\paragraph{Non linear decorrelation}

Two popular ICA algorithms based on non-linear decorrelation (hence
exploiting non Gaussianity) are JADE \cite{1999jade} and FastICA
\cite{1999fastica}.  In practice however, these algorithms do not
exactly solve an equation in the form~(\ref{eq:esteqng}).  Rather, for
algorithmic efficiency, they try to solve it under the additional
constraint that the components are uncorrelated \textit{i.e.}  that
condition~(\ref{eq:simplecorr}) is satisfied exactly.  The underlying
optimization engine is a joint diagonalization algorithm for JADE and
a fixed point technique for FastICA.


\paragraph{Localized decorrelation}

Efficient algorithms for solving the localized decorrelation
conditions~(\ref{eq:esteqloc}) are based on assuming some regularity
in the variance profiles: the sequences $\{\sigma_{ip}^2\}$ are
approximated as being constant over small domains.  Hence, the global
set $[1,\nbe]$ is partitioned into $Q$ subsets $\mathcal{I}_1,\ldots,
\mathcal{I}_Q$, each containing a number $\nbe_q$ of points (so that
$\nbe=\sum_{q=1}^Q \nbe_q$).  In practice, these pixel subsets are
(well chosen) spatial regions.
With a slight abuse of notation, we write
$\sigma_{ip}^2=\sigma_{iq}^2$ if $p\in\mathcal{I}_q$.
Then, a small amount of maths turns the decorrelation
conditions~(\ref{eq:esteqloc}) into
\begin{equation}
  \left[ 
    \sum_{q=1}^Q \nbe_q \Sig_q^{-1} \W \widehat \R_q \W\adj
  \right]_{ij}
  = 0
  \quad\mbox{for}\quad 1\leq i\neq j\leq\nbs
\end{equation}
where $\widehat\R_q$ is a localized covariance matrix 
\begin{equation}
  \widehat\R_q = \frac1{\nbe_q}\sum_{p\in q} \by(p)\by(p)\adj
  \quad\mbox{and}\quad
  \Sig_q = \mathrm{diag}(\sigma_{1q}^2, \ldots, \sigma_{\nbs q}^2)
  .
\end{equation}
An important point here is that by assuming piecewise constant
variance profiles, the localized decorrelation condition can be
expressed entirely in terms of the localized covariance matrices
$\widehat\R_q$.
Hence the localized covariance matrices appear as \emph{sufficient
  statistics} in this model.  Even better, the likelihood of $\A$ can
be understood as a mismatch between these statistics and their
predicted form, namely $\R_q=\A\Sig_q\A\adj$.  Specifically, in this
model the probability $p(\by(1),\ldots\by(P)| \A, \Sig)$ of the data
given $\A$ and the set $\Sig = \{ \Sig_1,\ldots,\Sig_Q \}$ of
covariance matrices is given by
\begin{displaymath}
  \log p(\by(1),\ldots\by(P)| \A, \Sig)  
  =
  - \phi(\A,\Sig) + \mathrm{cst}
\end{displaymath}
where function $\phi$ is defined as
\begin{equation}\label{eq:specmm}
  \phi(\A,\Sig) = \sum_q P_q K(\widehat \R_q, \A \Sig_q \A\adj)
\end{equation}
and where $K(\cdot, \cdot)$ is a measure of divergence between two
matrices defined as
\begin{equation}
  K(\R_1,\R_2) = 
  \frac12
  \left(
    \mathrm{trace}(\R_1\R_2\inv) -\log\det(\R_1\R_2\inv) -\nbs
  \right)
\end{equation}
This shows that maximum likelihood estimation of $\A$ amounts to the
minimization of the weighted mismatch~(\ref{eq:specmm}) between the
set of localized covariance matrices $\widehat\R_q$ (computed from the
data) and their expected value $\R_q=\A \Sig_q \A\adj$ (predicted by
the model).

In the noise-free case considered here, it turns out that there is a
simple and very efficient algorithm (due to D.T. Pham) for minimizing
the spectral mismatch.

\section{SMICA} \label{sec:smica}

We have developed a component separation technique dubbed
\textsc{SMICA} for `spectral matching \textsc{ICA}' which is based on
the ideas sketched at previous section but improves on them in several
ways.

In its simplest form, \textsc{SMICA} is based on \emph{spectral}
statistics, that is, on statistics which are localized not in space
but in frequency.  These statistics are binned auto- and cross-spectra
of the channels.  More specifically, for a given set of
$N_{\mathrm{chann}}$ multi-channels maps $\{ y_i(p) \}$, we form for
each $(\ell,m)$ the $N_{\mathrm{chann}}\times 1$ vector $\by(\ell, m)$
of their harmonic coefficients and define
\begin{displaymath}
  \widehat\R(\ell)
  =
  \frac1{2\ell+1}\sum_{m=-\ell}^{m=+\ell} \by(\ell, m)\by(\ell, m)\adj
\end{displaymath}
These empirical spectral covariance matrices are then binned.  In the
simplest case, we define $Q$ top-hat bins, with the $q$-th frequency
bin contains all frequencies $\ell$ between $\ell_q^{\mathrm{min}}$
and $\ell_q^{\mathrm{max}}$. We consider the binned spectra:
\begin{displaymath}
  \widehat\R_q = 
  \frac1{P_q}
  \sum
  _{\ell=\ell_q^{\mathrm{min}}}
  ^{\ell=\ell_q^{\mathrm{max}}}
  (2\ell+1)
  \widehat\R(\ell)
  \quad\mbox{where}\quad
  P_q
  =
  \sum
  _{\ell=\ell_q^{\mathrm{min}}}
  ^{\ell=\ell_q^{\mathrm{max}}}
  (2\ell+1)
  .
\end{displaymath}
Here $P_q$ is the number of Fourier modes summed together in a single
estimate $\widehat\R_q$.

The mixture model $\by=\A\bs+\bn$ predicts that the empirical spectra
$\R_q$ have an expected value
\begin{displaymath}
  \R_q = \langle\widehat \R_q \rangle = \A\Sig_q\A\adj + \N_q
\end{displaymath}
where $\Sig_q$ are the binned spectral covariance matrix for the
components in bin $q$ and $\N_q$ is the same for noise, assumed to be
uncorrelated from the components.
The unknown parameters can be collected in a big vector $\theta$:
\begin{displaymath}
  \theta = \left\{ \A, \{\Sig_q \}, \{ \N_q\}\right\}
\end{displaymath}
but in practice we will not fit such a large model.  Many constraints
can be imposed on $\theta$.  A typical choice is to assume that the
components are uncorrelated between themselves and that the noise also
is uncorrelated between channels.  Such a choice would result
in a smaller parameter set
\begin{displaymath}
  \theta = \left\{ \A, \{\mathrm{diag }\Sig_q \}, \{ \mathrm{diag}\N_q\}\right\}
\end{displaymath}
but infinitely many other options are possible, both more stringent
(like assuming that the noise in each channel is a smooth function of
the bin index $q$) or less stringent (like assuming that some
components may not be uncorrelated).
In the following, we do not assume a specific parametrization of the
binned spectral covariance matrices.  Rather, we denote
where $\theta$ is some parameter set which uniquely determines the
values of $\A$ and each $\R_q$ and $\N_q$:
\begin{displaymath}
  \{ \R_q \}
  =
  \{ \R_q (\theta) \}
  =
  \{  \A (\theta)\Sig_q (\theta)\A (\theta)\adj + \N_q  (\theta) \}
\end{displaymath}
\textsc{SMICA} determines the set $\theta$ of unknown parameters by
fitting the empirical spectral covariance matrices to whichever
structure is predicted by the model.  Specifically, the unknown
parameters are found by minimizing the ``spectral mismatch''
\begin{equation}\label{eq:smicamm}
  \phi(\theta) = \sum_q \nbe_q\, K(\widehat \R_q, \R_q(\theta) )
\end{equation}
averaged across bins.  Some comments are in order regarding the
matching criterion, the issue of non stationarity and practical
implementation.

\paragraph{Matching criterion}

The reason for choosing this particular form of mismatch between data
and model is that minimizing~(\ref{eq:smicamm}) is identical to
maximizing the likelihood of the data in a model where all components
are 1) Gaussian 2) stationary and 3) have harmonic spectra which are
constant over bins.  Of course, these assumptions are not met in
practice so one could choose a different criterion for matching
$\widehat\R_q$ to $\R_q(\theta)$ but we have little statistical
guidance for picking up an alternate matching measure.  Furthermore,
the assumptions 1) and 2) are met by the CMB and 3) is approximately
correct for narrow bins.  In addition, the failure of stationarity can
be alleviated by using localized statistics (see below).

\paragraph{Non stationarity and localization}

The spectral approach to building a likelihood function has some
benefits, in particular the fact that it is perfectly suited to
describing the statistical properties of the CMB.  
Another beneficial side effect is that it makes it easy to deal with
varying resolution from channel to channel as long as the beam can be
considered to be symmetrical

However, going straight away to harmonic space seems unreasonable to
deal with highly non stationary components such as the galactic
components.  This issue can be addressed to some extent by resorting
to localized spectral statistics.  It is a simple matter to use
spatial window functions to partition the sky into spatial domains \cite{2005cardoso}.
Although not a perfect solution, it certainly
allows to capture a good deal of the non stationary features of the
galactic sky.

\paragraph{Implementation}

The definition of the spectral matching criterion~(\ref{eq:smicamm})
encapsulates all of the statistical modeling but leaves open the
separate and possibly tricky issue of minimizing $\phi(\theta)$.\footnote{
We note in passing that some authors seem to make a confusion between the objective function (the criterion which has to be minimised, which derives from a statistical model) and the algorithm used for minimization. For instance, some authors use the terms ``EM method", or ``MCMC method", to design a method in which they use the EM algorithm, or Monte-Carlo Markov Chains. This is rather infortunate, and contributes to a certain level of confusion.}
Because the criterion is a likelihood in disguise, it is possible to
use the EM algorithm for its minimization, with the components taken
as latent variables.  However, EM is often not fast enough and also is
not able to deal with arbitrary parametrization of $\Sig_q(\theta)$
and $\N_q(\theta)$.  It has been found necessary to use general
optimization techniques.  A conjugate gradient algorithm can be
implemented because a reasonably tractable expression for the gradient
of the criterion is available as:
\begin{displaymath}
  \frac{\partial\phi}{\partial\theta}
  =
  \sum_q \nbe_q\, \mathrm{trace}
  \left(
    \R_q(\theta)\inv
    (\R_q(\theta)-\widehat\R_q) 
    \R_q(\theta)\inv
    \frac{\partial\R_q(\theta)}{\partial\theta}
  \right)
\end{displaymath}
However, in our context, the conjugate gradient algorithm also
requires preconditioning.  A preconditioner can be classically
obtained as the inverse of the Fisher information matrix
$\mathrm{FIM}(\theta)$ which is taken as an approximation to the
Hessian of $\phi(\theta)$:
\begin{displaymath}
  \frac{\partial^2\phi}{\partial\theta^2}
  \approx
  \mathrm{FIM}(\theta)
  =
  \sum_q \nbe_q\, \mathrm{trace}
  \left(
    \R_q(\theta)\inv
    \frac{\partial\R_q(\theta)}{\partial\theta}
    \R_q(\theta)\inv
    \frac{\partial\R_q(\theta)}{\partial\theta}
  \right)
\end{displaymath}

\paragraph{Mismatch control and error bars}

A benefit of the \textsc{SMICA} approach is that it comes with a
built-in measure of the quality of the model.  Indeed, if we properly
fit all the auto-cross spectra, then $\phi(\theta)$ should be
`statistically small'.  Visual control of the quality of the spectral
matching is obtained by plotting $\phi_q= \nbe_q\, K(\widehat \R_q,
\R_q(\hat\theta) )$ versus $q$ where $\hat\theta$ is the minimizer of
$\phi(\theta)$.  This quantity should be understood as a $\chi^2$.  If
the model holds (Gaussian stationary components and noise) and when
all spectral parameters are freely estimated $\phi_q$ behaves
approximately as a $\chi^2$ with a number of degrees of freedom equal
to $N_1-N_2$ where $N_1=N_{\mathrm{chann}}(N_{\mathrm{chann}}+1)/2$ is
the number of degrees of freedom for an sample covariance matrix of
size $N_{\mathrm{chann}}$ and where
$N_2=N_{\mathrm{comp}}+N_\mathrm{chann}$ is the number of adjustable
spectral parameters (the variances of each components and noise levels
in a given frequency bin).

\section{Other blind, semi-blind, or model learning methods}

This paper would not be complete without a quick review of some of the recent work. We quote here a few papers which we think deserve reading for further exploration of component separation issues and methods. Although unevenly mature, these methods provide complementary approaches, with advantages ad drawbacks which deserve to be investigated.

\subsection{FastICA}

A blind component separation based on the FastICA method has been
developed for CMB data reduction by Baccigalupi et al
\cite{2000MNRAS.318..769B}, with an extension to the full sky by Maino
et al \cite {2002MNRAS.334...53M}.  This blind approach uses, as
``engine'' for component separation, a measure of independence based
on non-Gaussianity.  Therefore, it is essentially equivalent to
finding components which cancel non-linear correlations in the sense
of equation \ref{eq:esteqng}.

For CMB applications, characterizing independence via non linear
correlations of the form~\ref{eq:esteqng} has some limitations.
Firstly, theory shows that this characterization allow for the
separation of at most one Gaussian component~\cite{Cardoso-ProcIEEE}.  The
Gaussian component is somehow found ``by default'', as the particular
component which is orthogonal to (uncorrelated with) all other non
Gaussian components.  This is a concern for component separation
performed with the CMB as the main target.
Secondly, the non-linear decorrelation conditions do not take the
noise into account.  Even though this can be fixed in some ad hoc
fashion, it is computationally demanding to do it in maximum
likelihood sense.
Finally, pixel space implementations cannot easily handle
channel-dependent beams (unless explicit beam deconvolution is
performed).  If, to circumvent this problem, one considers harmonic
space implementation, performance suffers from the fact that Fourier
tend to be more Gaussian than the original, pixel-domain maps.


FastICA, however, can outperform other component separation methods
for some applications.  Spectral based methods (like SMICA) cannot
blindly separate two components if their angular power spectra are
proportional.  FastICA does not suffer from this limitation and
therefore has an edge for separating galactic components. If all
galactic components have similar power spectra (say, proportional to
$\ell^{-3}$) then SMICA is expected to perform poorly without prior
information.

Although both FastICA and SMICA are blind methods entering in the
general class of ``independent component analysis'', it should thus be
stressed that they are conceptually very different.  Performance,
therefore, is expected to be very different also, and to depend on the
actual properties of the data sets.

FastICA has been used on COBE and on WMAP data \cite{2003MNRAS.344..544M,2007MNRAS.374.1207M}.

\subsection{Other recent developments}

A ``semi-blind" approach to component separation has been proposed by Tegmark and collaborators in a work where they model the foreground emissions using a number of physical parameters, which they estimate directly in the data sets \cite{2000ApJ...530..133T}. They estimate the impact of estimating these extra parameters in terms of accuracy loss on parameters of interest for CMB science. This paper was the first to address seriously the problem of component spectral indices varying over the sky.

Martinez-Gonzalez and collaborators have proposed a method for the extraction of the CMB specifically and for the estimation of its power spectrum \cite{2003MNRAS.345.1101M}. The EM algorithm is the main tool of the implementation. 

Eriksen and collaborators have developed a method based on a refined modeling of the astrophysical components, and fitting this model to the data to obtained estimates of foreground parameters \cite{2006ApJ...641..665E}. The fit of the parameters is made pixel by pixel at low-resolution using a MCMC techinque for exploring the likelihood. After this first ``model learning" step, the parameters obtained are used to estimate high resolution component maps.

Recently, Hansen and collaborators have proposed a CMB cleaning method based on a wavelet fit of component emissions obtained by differencing observations in different channels, and subtraction of the fit from observations made at frequencies where the CMB dominates \cite{2006ApJ...648..784H}.

Bonaldi and collaborators have recently published a paper  for estimating parameters of emission of astrophysical components (emission laws, described by spectral indices). The statistics used are based on estimations of the correlations of the observations using a subset of points on the sphere \cite{2006MNRAS.373..271B}. 

An alternate way of performing component separation has been proposed by Bobin and collaborators, based on sparse representations of the various emissions \cite{2006Bobin}. The basic principle of this method consists in decomposing the observations in a set of (redundant) dictionnaries chosen so that each component can be represented sparsely in one of the dictionnaries. Separation is achieved by minimizing the number of coefficients required to represent the data set.

A comparison of these different methods on a common data set, for investigating their strengths and weaknesses and evaluating their relative performance for various objectives would be an interesting work to improve the quality of component separation with the data set of upcoming space missions.

\section{Conclusion and prospects} \label{sec:conclusion}

With improving data quality and increasingly demanding performance in component characterisation, component separation will play an important role in the analysis of CMB data sets in the next decade.

In this paper, we have reviewed the main issues for component separation, concentrating on diffuse components specifically. 

Although substantial work has been performed, open questions remain. Polarisation, for instance, is one of the next major objectives of CMB science, for which much better sensitivities are required, and for which foreground emission is poorly known... Time varying sources, as the emission due to zodiacal light (modulated by the trajectory of the instrument in the ecliptic), as solar system objects in general, and as intrinsically time-varying radio sources, require specific methods tailored for their extraction.

The upcoming Planck data set, expected to become available to the Planck consortium in 2008, will provide a fantastic and challenging data set for extracting the emission from all astrophysical processes
emitting in the millimeter range. 

%
%
%
\input{referenc-v2}



\printindex
\end{document}

%% file: referenc-v2.tex
%
%

%
%

%% file: delabrouille-diffuse-nopdf.bbl
\begin{thebibliography}{99.}
%
%
%

\bibitem[Aghanim et al. (1996)]{1996A&A...311....1A} Aghanim, N., Desert, 
F.~X., Puget, J.~L., \& Gispert, R.\ 1996, \emph{Ionization by early quasars and cosmic microwave background anisotropies}, Astronomy and Astrophysics, 311, 1 

\bibitem[Audit \& Simmons (1999)]{1999MNRAS.305L..27A} Audit, E., \& 
Simmons, J.~F.~L.\ 1999, 
\emph{The kinematic Sunyaev-Zel'dovich effect and transverse cluster velocities},
MNRAS, 305, L27 

\bibitem[Baccigalupi et al.(2000)]{2000MNRAS.318..769B} Baccigalupi, C., et 
al.\ 2000, 
\emph{Neural networks and the separation of cosmic microwave background and astrophysical signals in sky maps}, MNRAS, 318, 769 

\bibitem[Barreiro (2005)]{barreiro-compact-sources} Barreiro, R.~B.\ 2006, 
\emph{Techniques for compact source extraction on CMB maps}, to appear in ``Data Analysis in Cosmology", Springer-Verlag Lecture Notes in Physics. Valencia, 6-10 September 2004 (ArXiv Astrophysics e-prints, arXiv:astro-ph/0512538) 

\bibitem[Barreiro et al. (2004)]{2004MNRAS.351..515B} Barreiro, R.~B., 
Hobson, M.~P., Banday, A.~J., Lasenby, A.~N., Stolyarov, V., Vielva, P., \& 
G{\'o}rski, K.~M.\ 2004, \emph{Foreground separation using a flexible maximum-entropy algorithm: an application to COBE data}, MNRAS, 351, 515 

\bibitem[Bennett et al. (2003)]{2003ApJS..148....1B} Bennett, C.~L., et al. 
2003, \emph{First-Year Wilkinson Microwave Anisotropy Probe (WMAP) Observations: Preliminary Maps and Basic Results}, ApJ Supplement Series, 148, 1 

\bibitem[Bennett et al. (2003)]{2003ApJS..148...97B} Bennett, C.~L., et al. 2003, 
\emph{First year Wilkinson Microwave Anisotropy Probe (WMAP) observations: foreground emission}, ApJ Supplement Series, 148, 97 

\bibitem[Beno{\^i}t et al. (2004)]{2004A&A...424..571B} Beno{\^i}t, A., et 
al.\ 2004, \emph{First detection of polarization of the submillimetre diffuse galactic dust emission by Archeops}, Astronomy and Astrophysics, 424, 571 

\bibitem[Birkinshaw (1999)]{1999PhR...310...97B} Birkinshaw, M.\ 1999, 
\emph{The Sunyaev Zel'dovich effect}
Physics Reports, 310, 97 

\bibitem[Bobin (2006)]{2006Bobin} Bobin, J., Moudden, Y., Starck, J.-L. and Elad, M.\ 2006, 
\emph{Morphological Diversity and Source Separation}
IEEE Transaction on Signal Processing, in press.

\bibitem[Bonaldi et al.(2006)]{2006MNRAS.373..271B} Bonaldi, A., Bedini, 
L., Salerno, E., Baccigalupi, C., \& de Zotti, G.\ 2006, \emph{Estimating the spectral indices of correlated astrophysical foregrounds by a second-order statistical approach}, MNRAS, 373, 271 

\bibitem[Bouchet \& Gispert (1999)]{1999NewA....4..443B} Bouchet, F.~R., \& 
Gispert, R.\ 1999, \emph{Foregrounds and CMB experiments I. Semi-analytical estimates of contamination}, New Astronomy, 4, 443 

\bibitem[Cardoso (1998)]{Cardoso-ProcIEEE} Cardoso, J.-F.\ 1998, \emph{Blind signal separation: statistical principles}, Proceedings of the IEEE. Special issue on blind identification and estimation, vol. 9, no 10, pp.~2009-2025

\bibitem[Cardoso (1999)]{1999jade} Cardoso, J.-F.\ 1999, \emph{High-order contrasts for independent component analysis}, Neural Computation, vol. 11, no 1, pp.~157-192

\bibitem[Cardoso et al. (2005)]{2005cardoso} Cardoso, J.-F., et al. \ 2005, \emph{Statistiques direction-multip\^ole pour la s\'eparation de composantes dans le fonds de rayonnement cosmologique}, Actes du GRETSI, Louvain-la-Neuve, Belgique

\bibitem[de Bernardis et al. (2000)]{2000Natur.404..955D} de Bernardis, P., 
et al.\ 2000, \emph{A flat Universe from high-resolution maps of the cosmic microwave background radiation}, Nature, 404, 955 

\bibitem[Delabrouille et al. (2003)]{Del2003} Delabrouille,~J., Cardoso~J.-F. and Patanchon,~G.\ 2003,
{\emph {Multi-detector multi-component spectral matching and application for CMB data analysis}}, 
 MNRAS, vol.~346, no~4,
pp.~1089-1102

\bibitem[Dickinson et al. (2004)]{2004MNRAS.353..732D} Dickinson, C., et 
al.\ 2004, \emph{High-sensitivity measurements of the cosmic microwave background power spectrum with the extended Very Small Array}, MNRAS, 353, 732 

\bibitem[Dickinson et al. (2006)]{2006ApJ...643L.111D} Dickinson, C., 
Casassus, S., Pineda, J.~L., Pearson, T.~J., Readhead, A.~C.~S., \& Davies, 
R.~D.\ 2006,
\emph{An Upper Limit on Anomalous Dust Emission at 31 GHz in the Diffuse Cloud [LPH96] 201.663+1.643},
Astrophysical Journal Letters, 643, L111 
 
\bibitem[Draine \& Lazarian (1998)]{1998ApJ...494L..19D} Draine, B.~T., \& 
Lazarian, A.\ 1998, \emph{Diffuse Galactic Emission from Spinning Dust Grains},
Astrophysical Journal Letters, 494, L19 

\bibitem[Eriksen et al. (2004)]{2004ApJ...612..633E} Eriksen, H.~K., Banday, 
A.~J., G{\'o}rski, K.~M., \& Lilje, P.~B.\ 2004, {\emph {On foreground removal from the WMAP data by an ILC method: limitations and implications}}, ApJ, 612, 633

\bibitem[Eriksen et al. (2006)]{2006ApJ...641..665E} Eriksen, H.~K., et al.\ 
2006, \emph{Cosmic Microwave Background Component Separation by Parameter Estimation}, ApJ, 641, 665 

\bibitem[Fern{\'a}ndez-Cerezo et al. (2006)]{2006MNRAS.370...15F} 
Fern{\'a}ndez-Cerezo, S., et al.\ 2006, 
\emph{Observations of the cosmic microwave background and galactic foregrounds at 12-17GHz with the COSMOSOMAS experiment},
MNRAS, 370, 15 

\bibitem[Gruzinov \& Hu (1998)]{1998ApJ...508..435G} Gruzinov, A., \& Hu, 
W.\ 1998, \emph{Secondary Cosmic Microwave Background Anisotropies in a Universe Reionized in Patches}, The Astrophysical Journal, 508, 435 

\bibitem[Hansen et al.(2006)]{2006ApJ...648..784H} Hansen, F.~K., Banday, 
A.~J., Eriksen, H.~K., G{\'o}rski, K.~M., \& Lilje, P.~B.\ 2006, \emph{Foreground Subtraction of Cosmic Microwave Background Maps Using WI-FIT (Wavelet-Based High-Resolution Fitting of Internal Templates)}, ApJ, 648, 
784 

\bibitem[Haslam et al. (1981)]{1981A&A...100..209H} Haslam, C.~G.~T., Klein, 
U., Salter, C.~J., Stoffel, H., Wilson, W.~E., Cleary, M.~N., Cooke, D.~J., 
\& Thomasson, P.\ 1981, 
\emph{A 408 MHz all-sky continuum survey. I - Observations at southern declinations and for the North Polar region}, Astronomy and Astrophysics, 100, 209 

\bibitem[Hinshaw et al. (2006)]{2006astro.ph..3451H} Hinshaw, G., et al.\ 
2006, \emph{Three-Year Wilkinson Microwave Anisotropy Probe (WMAP) Observations: Temperature Analysis}, ArXiv Astrophysics e-prints, arXiv:astro-ph/0603451 

\bibitem[Hobson \& Lasenby (1998)]{1998MNRAS.298..905H} Hobson, M.~P., \& 
Lasenby, A.~N.\ 1998, \emph{The entropic prior for distributions with positive and negative values}, MNRAS, 298, 905 

\bibitem[Hobson et al. (1998)]{1998MNRAS.300....1H} Hobson, M.~P., Jones, 
A.~W., Lasenby, A.~N., \& Bouchet, F.~R.\ 1998, \emph{Foreground separation methods for satellite observations of the cosmic microwave background}, MNRAS, 300, 1 

\bibitem[Hu \& Dodelson (2002)]{2002ARA&A..40..171H} Hu, W., \& Dodelson, 
S.\ 2002,  {\emph {Cosmic Microwave Background Anisotropies}}, Annual Review of Astronomy and Astrophysics, 40, 171 

\bibitem[Hyvärinen (1999)]{1999fastica} Hyvvärinen, A.\ 1999, {\emph{Fast and Robust Fixed-Point Algorithms for Independent Component Analysis}}, IEEE Transactions on Neural Networks 10(3):626-634

\bibitem[Jones et al. (2005)]{2005astro.ph..7494J} Jones, W.~C., et al.\ 
2005, \emph{A Measurement of the Angular Power Spectrum of the CMB Temperature Anisotropy from the 2003 Flight of Boomerang}, ArXiv Astrophysics e-prints, arXiv:astro-ph/0507494 

\bibitem[Kogut et al. (1996)]{1996ApJ...460....1K} Kogut, A., Banday, A.~J., 
Bennett, C.~L., Gorski, K.~M., Hinshaw, G., \& Reach, W.~T.\ 1996,
\emph{High-Latitude Galactic Emission in the COBE Differential Microwave Radiometer 2 Year Sky Maps}, 
The Astrophysical Journal, 460, 1 

\bibitem[Komatsu et al. (2003)]{2003ApJS..148..119K} Komatsu, E., et al.\ 
2003, \emph{First-Year Wilkinson Microwave Anisotropy Probe (WMAP) Observations: Tests of Gaussianity}, ApJ Supplement Series, 148, 119 

\bibitem[Kuo et al. (2004)]{2004ApJ...600...32K} Kuo, C.~L., et al.\ 2004, \emph{High-Resolution Observations of the Cosmic Microwave Background Power Spectrum with ACBAR}, The Astrophysical Journal, 600, 32 

\bibitem[Lamarre et al. (2000)]{2000ApL&C..37..161L} Lamarre, J.~M., et al.\ 
2000, \emph{The High Frequency Instrument of Planck: Design and Performances}, Astrophysical Letters Communications, 37, 161 

\bibitem[Lamarre et al. (2003)]{2003NewAR..47.1017L} Lamarre, J.~M., et al.\ 2003, 
\emph{The Planck High Frequency Instrument, a third generation CMB experiment, and a full sky submillimeter survey},
New Astronomy Review, 47, 1017 

\bibitem[Leitch et al. (1997)]{1997ApJ...486L..23L} Leitch, E.~M., Readhead, 
A.~C.~S., Pearson, T.~J., \& Myers, S.~T.\ 1997, 
\emph{An Anomalous Component of Galactic Emission},
Astrophysical Journal Letters, 486, L23 

\bibitem[Maino et al.(2002)]{2002MNRAS.334...53M} Maino, D., et al.\ 2002, 
\emph{All-sky astrophysical component separation with Fast Independent Component Analysis (FASTICA)}, MNRAS, 334, 53 

\bibitem[Maino et al.(2003)]{2003MNRAS.344..544M} Maino, D., Banday, A.~J., 
Baccigalupi, C., Perrotta, F., \& G{\'o}rski, K.~M.\ 2003, 
\emph{Astrophysical component separation of COBE-DMR 4-yr data with FASTICA}, MNRAS, 344, 544 

\bibitem[Maino et al.(2007)]{2007MNRAS.374.1207M} Maino, D., Donzelli, S., 
Banday, A.~J., Stivoli, F., \& Baccigalupi, C.\ 2007, 
\emph{Cosmic microwave background signal in Wilkinson Microwave Anisotropy Probe three-year data with FASTICA}
MNRAS, 374, 1207 

\bibitem[Mandolesi et al. (2000)]{2000ApL&C..37..151M} Mandolesi, N., 
Bersanelli, M., Burigana, C., \& Villa, F.\ 2000, \emph{The Planck Low Frequency Instrument}, Astrophysical Letters Communications, 37, 151 

\bibitem[Maris et al. (2006)]{2006A&A...452..685M} Maris, M., Burigana, C., 
\& Fogliani, S.\ 2006, Astronomy and Astrophysics, 452, 685 

\bibitem[Mart{\'{\i}}nez-Gonz{\'a}lez et al.(2003)]{2003MNRAS.345.1101M} 
Mart{\'{\i}}nez-Gonz{\'a}lez, E., Diego, J.~M., Vielva, P., \& Silk, J.\ 
2003, \emph{Cosmic microwave background power spectrum estimation and map reconstruction with the expectation-maximization algorithm}, MNRAS, 345, 1101 

\bibitem[McCullough et al. (1999)]{1999ASPC..181..253M} McCullough, 
P.~R., \& et al.\ 1999, 
\emph{Implications of Halpha Observations for Studies of the CMB},
ASP Conf.~Ser.~181: Microwave Foregrounds, 181, 253 

\bibitem[Moudden et al. (2005)]{2005moudden}
Moudden, Y., Cardoso, J.-F., Starck, J.-L., Delabrouille, J.\ 2005, \emph{Blind Component Separation in Wavelet Space: Application to CMB Analysis}, Eurasip Journal on Applied Signal Processing , 2005, 15 pp 2437-2454

\bibitem[Netterfield et al. (2002)]{2002ApJ...571..604N} Netterfield, C.~B., 
et al.\ 2002, \emph{A Measurement by BOOMERANG of Multiple Peaks in the Angular Power Spectrum of the Cosmic Microwave Background}, The Astrophysical Journal, 571, 604 

\bibitem[Patanchon et al. (2005)]{2005MNRAS.364.1185P} Patanchon, G., 
Cardoso, J.-F., Delabrouille, J., \& Vielva, P.\ 2005, \emph{Cosmic microwave background and foregrounds in Wilkinson Microwave Anisotropy Probe first-year data}, MNRAS, 364, 1185 

\bibitem[Ponthieu et al. (2005)]{2005A&A...444..327P} Ponthieu, N., et al.\ 
2005, \emph{Temperature and polarization angular power spectra of Galactic dust radiation at 353 GHz as measured by Archeops}, Astronomy and Astrophysics, 444, 327 

\bibitem[Puget et al. (1996)]{1996A&A...308L...5P} Puget, J.-L., Abergel, 
A., Bernard, J.-P., Boulanger, F., Burton, W.~B., Desert, F.-X., \& 
Hartmann, D.\ 1996, \emph{Tentative detection of a cosmic far-infrared background with COBE}, Astronomy and Astrophysics, 308, L5 

\bibitem[Readhead et al. (2004)]{2004ApJ...609..498R} Readhead, A.~C.~S., et 
al.\ 2004, \emph{Extended Mosaic Observations with the Cosmic Background Imager}, The Astrophysical Journal, 609, 498 

\bibitem[Rephaeli (2002)]{2002AIPC..616..309R} Rephaeli, Y.\ 2002, 
\emph{The Sunyaev-Zeldovich effect: Recent progress and future prospects}, 
AIP Conf.~Proc.~616: Experimental Cosmology at Millimetre Wavelengths, 616, 309 

\bibitem[Sazonov \& Sunyaev (1999)]{1999MNRAS.310..765S} Sazonov, S.~Y., \& 
Sunyaev, R.~A.\ 1999, 
\emph{Microwave polarization in the direction of galaxy clusters induced by the CMB quadrupole anisotropy}
MNRAS, 310, 765 

\bibitem[Shannon (1948)]{shannon-entropy} Shannon,\ 1948, 
\emph{A Mathematical Theory of Communication}
Bell System Technical Journal, vol. 27, pp. 379-423, 623-656, July, October, 1948
 
\bibitem[Smoot (1998)]{1998astro.ph..1121S} Smoot, G.~F.\ 1998, 
emph{Galactic Free-Free and H-alpha emission}, 
ArXiv Astrophysics e-prints, arXiv:astro-ph/9801121 

\bibitem[Stolyarov et al. (2002)]{2002MNRAS.336...97S} Stolyarov, V., 
Hobson, M.~P., Ashdown, M.~A.~J., \& Lasenby, A.~N.\ 2002, \emph{All-sky component separation for the Planck mission},  MNRAS, 336, 97 

\bibitem[Sunyaev \& Zeldovich (1972)]{1972CoASP...4..173S} Sunyaev, R.~A., 
\& Zeldovich, Y.~B.\ 1972, 
\emph{The Observations of Relic Radiation as a Test of the Nature of X-Ray Radiation from the Clusters of Galaxies}
Comments on Astrophysics and Space Physics, 4, 
173 

\bibitem[Sunyaev \& Zeldovich (1980)]{1980MNRAS.190..413S} Sunyaev, R.~A., 
\& Zeldovich, I.~B.\ 1980, 
\emph{The velocity of clusters of galaxies relative to the microwave background - The possibility of its measurement}
MNRAS, 190, 413 

\bibitem[Tegmark \& Efstathiou (1996)]{1996MNRAS.281.1297T} Tegmark, M., \& 
Efstathiou, G.\ 1996, \emph{A method for subtracting foregrounds from multifrequency CMB sky maps}, MNRAS, 281, 1297 

\bibitem[Tegmark et al.(2000)]{2000ApJ...530..133T} Tegmark, M., 
Eisenstein, D.~J., Hu, W., \& de Oliveira-Costa, A.\ 2000, 
\emph{Foregrounds and Forecasts for the Cosmic Microwave Background}, The Astrophysical Journal, 530, 133 

\bibitem[Tegmark et al. (2003)]{2003PhRvD..68l3523T} Tegmark, M., de 
Oliveira-Costa, A., \& Hamilton, A.~J.\ 2003, \emph{High resolution foreground cleaned CMB map from WMAP} Phys. Rev. D., 68, 123523 

\bibitem[Tristram et al. (2005)]{2005A&A...436..785T} Tristram, M., et al.\ 
2005, \emph{The CMB temperature power spectrum from an improved analysis of the Archeops data}, Astronomy and Astrophysics, 436, 785 

\bibitem[Valls-Gabaud (1998)]{1998PASA...15..111V} Valls-Gabaud, D.\ 1998, 
\emph{Cosmological applications of H-alpha surveys},
Publications of the Astronomical Society of Australia, 15, 111 

\bibitem[White \& Cohn (2002)]{2002AmJPh..70..106W} White, M., \& Cohn, 
J.~D.\ 2002, \emph{The theory of anisotropies in the cosmic microwave background}, American Journal of Physics, 70, 106 

\bibitem[Wiener (1949)]{1949WienerFilter} Wiener, N.\ 1949, \emph{Extrapolation, Interpolation, and Smoothing of Stationary Time Series}. New York: Wiley. ISBN 0262730057

\bibitem[Yamada et al. (1999)]{1999ApJ...522...66Y} Yamada, M., Sugiyama, 
N., \& Silk, J.\ 1999, \emph{The Sunyaev-Zeldovich Effect by Cocoons of Radio Galaxies}, The Astrophysical Journal, 522, 66 








\end{thebibliography}
